\documentclass[aip, pof, reprint]{revtex4-1}

\usepackage{subfig}
\usepackage{SIunits}
\usepackage{amsmath}
\usepackage{color}

\usepackage{graphicx}
\usepackage{float}
\usepackage{booktabs}
\usepackage{tabularx}
\begin{document}

%\preprint{}

\title{Combined free-stream disturbance measurements and receptivity studies in hypersonic wind tunnels by means of a slender wedge probe and DNS} %Title of paper



\author{Alexander Wagner}
\email[]{Alexander.Wagner@dlr.de}

\author{Erich Sch\"ulein}
\email[]{Erich.Sch\"ulein@dlr.de}

\author{Ren\'{e} Petervari}

\author{Klaus Hannemann}
\email[]{Klaus.Hannemann@dlr.de}
%\thanks{}
\affiliation{German Aerospace Center (DLR), Institute of Aerodynamics and Flow Technology, Bunsenstra{\ss}e 10, 37073 G\"ottingen, Germany}

\author{Syed R. C. Ali}
\email[]{S.Ali@tu-bs.de}
\affiliation{TU Braunschweig, Institute of Fluid Mechanics, Herman-Blenk-Str. 37, 38108 Braunschweig, Germany}

\author{Adriano Cerminara}
\email[]{A.Cerminara@soton.ac.uk}

\author{Neil D. Sandham}
\email[]{N.Sandham@soton.ac.uk}
%\thanks{}
\affiliation{University of Southampton, Aerodynamics and Flight Mechanics Research Group, Southampton SO17 1BJ, UK}


\date{\today}

\begin{abstract}

Combined free-stream disturbance measurements and receptivity studies in hypersonic wind tunnels were conducted by means of a slender wedge probe and direct numerical simulation. The study comprises comparative tunnel noise measurements at Mach 3, 6 and 7.4 in two Ludwieg tube facilities and a shock tunnel. Surface pressure fluctuations were measured over a wide range of frequencies and test conditions including harsh test environments not accessible to measurement techniques such as pitot probes and hot-wire anemometry. A good agreement was found between normalized pitot pressure fluctuations converted into normalized static pressure fluctuations and the wedge probe readings. Quantitative results of the tunnel noise are provided in frequency ranges relevant for hypersonic boundary layer transition. In combination with the experimental studies, direct numerical simulations of the leading-edge receptivity to fast and slow acoustic waves were performed for the slender wedge probe at conditions corresponding to the experimental free-stream conditions. The receptivity to fast acoustic waves was found to be characterized by an early amplification of the induced fast mode. For slow acoustic waves an initial decay was found close to the leading edge. At all Mach numbers, and for all considered frequencies, the leading-edge receptivity to fast acoustic waves was found to be higher than the receptivity to slow acoustic waves. Further, the effect of inclination angles of the acoustic wave with respect to the flow direction was investigated. An inclination angle was found to increase the response on the wave-facing surface of the probe and decrease the response on the opposite surface for fast acoustic waves. A frequency-dependent response was found for slow acoustic waves. 
The combined numerical and experimental approach in the present study confirmed the previous suggestion that the slow acoustic wave is the dominant acoustic mode in noisy hypersonic wind tunnels.
%The surface-to-free-stream transfer functions at the exact transducer locations were computed to estimate the free-stream disturbance levels based on the experimental results and the case considering solely slow acoustic waves or solely fast acoustic waves to be present in the free-stream. Good agreement was found with recent literature for the Mach 3 case.
\end{abstract}

\pacs{}% insert suggested PACS numbers in braces on next line
\maketitle %\maketitle must follow title, authors, abstract and \pacs

\section{Introduction}
\label{Introduction}

Free-stream disturbances are known to play an important role in the boundary layer transition process. An understanding of the transition process improved in supersonic and hypersonic boundary layers is in turn crucial for the design of vehicles operating at such speeds. Uncertainties in the transition prediction directly lead to uncertainties in the estimated viscous drag and the surface heat flux, which are both essential design parameters for vehicles operating in the hypersonic flow regime. 

Previous studies revealed that breakdown mechanisms are initial condition dependent.\cite{Singer1989} Free-stream disturbances, such as vorticity, sound, entropy inhomogeneity and microscale and macroscale particulates, enter the boundary layer as unsteady fluctuations of the basic state. This process is called receptivity.\cite{Morkovin1969} It establishes the initial conditions of the disturbance amplitude, frequency and phase of the breakdown of laminar flow.\cite{Saric2002a,Fedorov2003a} Therefore, the characterization of free-stream disturbances in the relevant frequency range and an understanding of how disturbances are entrained into the boundary layer are key aspects of studying boundary layer transition.\cite{Fedorov2010} Since the majority of the transition studies are conducted in noisy facilities it is of importance to determine the disturbance environment to correctly interpret the experimental results. Extensive reviews of the effect of tunnel noise on high speed boundary layer transition were conducted by Schneider\cite{Schneider2001a,Schneider2004}. 

In recent years considerable effort has been undertaken to characterize hypersonic wind tunnel noise world wide. The present study was motivated by numerous transition studies in shock tunnels for instance in T5 \cite{Adam1997,Germain1997,Rasheed2002,Parziale2012}, HIEST \cite{Tanno2009,Tanno2010,Nagayama2016} and HEG \cite{Wartemann2013a,Wagner2013a,Laurence2014,Ozawa2014,Laurence2016} and the lack of suitable techniques to investigate the disturbance environment in such tunnels. A number of experimental techniques are commonly used for this purpose in hypersonic wind tunnels. For instance hot wire anemometry (HWA) is widely used to quantify disturbances radiated from a supersonic turbulent boundary layer or to determine the source and the nature of the disturbances.\cite{Laufer1964, Anders1977} Recently, the technique was used by Masutti et al. \cite{Masutti2012} to characterize the disturbance level of the Mach 6 blow-down facility H3 at VKI. Unfortunately, the technique is not applicable to short-time impulse facilities such as high-enthalpy shock tunnels. Due to the limited bandwidth, approximately \unit{100}{kHz}, the high frequency content in such flows cannot be assessed. Furthermore, the total temperatures in such facilities are very high compared to blow-down or Ludwieg tube facilities. This reduces the achievable overheat ratio and thus thwarts the data reduction strategy introduced by Smits et al. \cite{Smits1983}. Furthermore, the harsh test environment and the impulsive nature of the flow most likely compromise the delicate HWA wires. Another popular technique widely used to assess free-stream disturbances is the pitot probe.\cite{Wendt1995,Bounitch2011,Rufer2012,Gromyko2013,Grossir2013,Mai2014} Although the technique is easy to realize, it suffers from a number of drawbacks. For instance, to avoid protective cavities, which lead to frequency-dependent damping effects and resonances, the transducers need to be flush-mounted facing the stagnation conditions. This puts the transducers at risk of excessive thermal loading and particulate impact, especially in shock tunnels. Furthermore, Chaudhry et al. \cite{Chaudhry2016} studied the transfer function of various pitot probe geometries, considering fast acoustic, slow acoustic and entropy disturbances. The transfer functions were found to be a strong function of the shock stand-off distance and the probe geometry, which is not standardized in shape and size, and thus makes the comparison of results obtained with different probes difficult.\cite{Heitmann2008}
A promising alternative to intrusive techniques is the non-intrusive focused laser differential interferometer technique applied by Parziale et al. \cite{Parziale2014} to conduct quantitative measures of density fluctuations in the reflected shock tunnel T5. The technique was first described by Smeets et al. \cite{Smeets1972,Smeets1973} and exhibits a very high frequency response (\unit{>10}{MHz}) and an adequate spatial resolution. The technique is limited to density fluctuations and unfortunately cannot easily be transferred between different facilities due to its elaborate setup. 

Recently, Tsyryulnikov et al. \cite{Tsyryulnikov2016} introduced a method of mode decomposition based on long-wave free-stream disturbance interactions with oblique shock waves at different shock wave angles. The method was applied in a hotshot wind tunnel using flat plate probes at different angles of attack. However, the method covers a limited frequency range of solely \unit{2-20}{kHz}. Ali et al. \cite{Ali2014} investigated the free-stream disturbance spectra in a Mach 6 wind tunnel by means of a cone probe in combination with HWA and a pitot probe. The experimental activities were complemented by a numerical study conducted by Schilden et al. \cite{Schilden2016}. The combined study also aimed for decomposing the measured free-stream disturbances into the three disturbance modes as introduced by Kovasznay \cite{Kovasznay1953}. For the investigated test cases the acoustic mode was found to be about one order of magnitude higher compared to the entropy mode whereas the vorticity mode was found to be negligible, which is in line with Pate's observation \cite{Pate1980}. \\

%% adriano text
Due to the difficulties in measuring the free-stream disturbance field in hypersonic wind tunnels with the current experimental techniques, a combined experimental-numerical approach is useful to characterize the environmental noise. This strategy takes advantage of the numerical capabilities of existing high-resolution direct numerical simulation (DNS) codes to accurately describe the physical mechanisms of boundary-layer receptivity from a prescribed disturbance field. In particular, a cross-validation process between the experimental data of a certain test conducted under particular wind-tunnel conditions, and a set of numerical results with the same test conditions and different imposed models of the free-stream disturbances, would allow a better evaluation of the sensitivity of the experimental data. At the same time, it allows an assessment of the disturbance-field model (among those considered in the numerical simulations) that provides the closest receptivity characteristics to the experimental results. According to Schneider \cite{Schneider2015} this provides a basis for a gradual calibration process of the numerical disturbance field aiming for a high-fidelity reconstruction of the main characteristics of the environmental noise for a particular hypersonic wind tunnel. The main challenge related to the numerical simulations is to assume an initial disturbance field which already contains the most relevant information concerning the actual free-stream noise, in terms of, e.g. disturbance type, amplitude, orientation, frequencies, etc. Despite its high complexity, some simplifications of the disturbance field can be made, according to numerical and theoretical studies available in the literature. For example, in a recent numerical study, Duan et al. \cite{Duan2014a} showed that the noise generated by a fully turbulent boundary layer in a flow at Mach 2.5 over a flat plate is mainly characterized by acoustic disturbances with wavefront orientation and phase speed belonging to the class of slow acoustic waves. This is an indication that slow acoustic modes are efficiently produced by turbulent boundary layers on the wind tunnel walls. Moreover, a theoretical study of McKenzie and Westphal \cite{McKenzie1968} showed that incident entropy/vorticity waves can generate intense acoustic waves behind the oblique shock, which was confirmed by the numerical study of Ma and Zhong \cite{Ma2005} on the receptivity of a Mach 4.5 flow over a flat plate. The latter study showed that the boundary-layer disturbances are mostly induced by fast acoustic waves generated behind the shock resulting from the interaction of free-stream entropy and vorticity waves. This demonstrates that, also in the presence of non-acoustic wave types, the acoustic waves turn out to be the most influential disturbances in the boundary-layer receptivity process, due to the fundamental role of the shock in establishing the post-shock wave structure. This motivates the present numerical study of the effects of slow and fast acoustic waves on the receptivity mechanism.

The present study contributes to the broad effort taken to characterize free-stream disturbances in hypersonic wind tunnels by introducing a wedge-shaped probe designed to measure free-stream disturbances over a wide frequency range in hypersonic wind tunnels and in particular in hypersonic shock tunnels with harsh test environments. The wedge probe shape provides a number of advantages over a conical shape when it comes to its practical application. However, it follows the same principle of using a slender probe instead of a blunt geometry to avoid strong shocks and thus the complexity of a subsonic flow field associated with complex amplification of the tunnel disturbances.\cite{McKenzie1968,Anyiwo1982} The probe was successfully used in three hypersonic wind tunnels, the DLR High Enthalpy Shock Tunnel G\"ottingen (HEG), the DNW Ludwieg tube (RWG) and the TU Braunschweig Ludwieg tube (HLB), covering Mach 3, 6 and 7.4.\cite{Wagner2016a} Information on the free-stream fluctuations are provided beyond the commonly used root mean square (RMS) of pressure readings. Instead, free-stream disturbance amplitudes are provided in specific frequency ranges, for instance those relevant to second mode dominated transition. Since tunnel noise is known to be dominated by acoustic modes mostly radiated at an angle from the nozzle wall boundary layer, the present paper concentrates on a numerical investigation of the role of fast and slow acoustic modes with different angles of incidence, as the two relevant modes for boundary layer transition.\cite{Fedorov2001a} 
In the scope of the numerical investigation, two-dimensional simulations were performed on a cylinder-wedge geometry representing the introduced probe, inserting planar fast and slow acoustic waves with multiple frequencies and random phase as free-stream disturbances. The obtained numerical transfer functions were combined with the experimental results in order to provide an estimation of the noise levels in the hypersonic wind tunnels in which the experiments were performed.

\section{Experimental Setup}
\label{sec:ExperimentalSetup}
\subsection{Hypersonic Test Facilities}

The initial tests of the present study were conducted in the DLR High Enthalpy Shock Tunnel G\"ottingen at Mach 7.4 and the DNW-RWG Ludwieg Tube in G\"ottingen at Mach 6 and Mach 3. Subsequently, a series of tests was conducted in the Hypersonic Ludwieg Tube (HLB) of University of Braunschweig. The present section provides a brief overview of the main characteristics of each hypersonic wind tunnel, focusing on the mode of operation and the test conditions applied in the present study.

\subsubsection{The High Enthalphy Shock Tunnel G\"ottingen}

The High Enthalpy Shock Tunnel G\"ottingen (HEG) is a free-piston driven reflected shock tunnel providing a pulse of gas to a hypersonic nozzle at stagnation pressures of up to \unit{200}{MPa} and stagnation enthalpies of up to \unit{25}{MJ/kg}.\cite{Eitelberg1992,Eitelberg1994,Hannemann2008} Originally, the facility was designed to investigate the influence of high-temperature effects such as chemical and thermal relaxation on the aerothermodynamics of entry or reentry space vehicles. Since its first commissioning, the range of operating conditions was extended to allow investigations of the flow past hypersonic flight configurations from low altitude Mach 6 up to Mach 10 at approximately 33 km altitude.\cite{Hannemann2008} The overall length and mass of the facility is $\unit{60}{m}$ and $\unit{250}{t}$, respectively. 
	\begin{figure}[htbp]
		\centering
			\includegraphics[width=0.75\columnwidth]{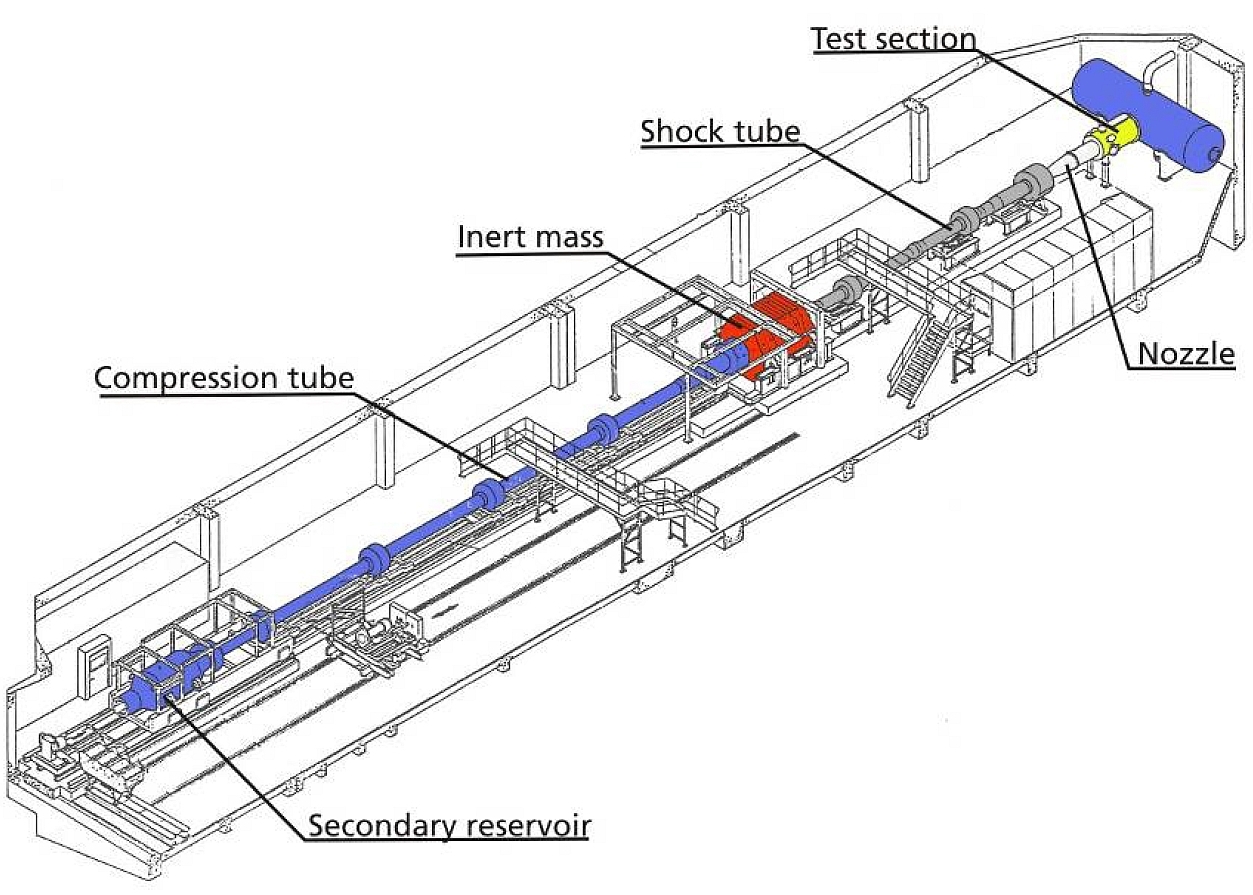}
		\caption{Schematic view of HEG, Martinez Schramm \cite{Schramm2008}.}
		\label{fig:HEG_color}
	\end{figure}
As shown in figure \ref{fig:HEG_color} the tunnel consists of three main sections. The driver section consists of a secondary reservoir which can be pressurized up to $\unit{23}{MPa}$ and a $\unit{33}{m}$ long compression tube. The adjoining shock tube (or driven tube) has a length of $\unit{17}{m}$. The shock tube is separated from the compression tube by a $\unit{3\,-\,18}{mm}$ stainless steel main diaphragm. The third section is separated by a thin diaphragm and consists of the Laval nozzle, the test section and the dump tank. For a test in HEG, pressurized air in the secondary reservoir is used to accelerate the piston down the compression tube. The driver gas in the compression tube is compressed quasi-adiabatically. When the burst pressure is reached, the main diaphragm ruptures and hot high-pressure gas expands into the shock tube. The shock wave produced is reflected at the end wall and provides the high-pressure, high-temperature gas that is expanded through a contoured convergent-divergent hypersonic nozzle after secondary diaphragm rupture. The nozzle exit diameter is \unit{0.59}{m} and the expansion ratio 218. In the scope of the present article, HEG was operated at the conditions listed in Table \ref{tab:HEGconditions}. The free-stream probe was positioned on the nozzle axis about \unit{100}{mm} downstream the nozzle exit.

\begin{table}[hbtp]
	\centering
		\begin{tabular}{l|c|c|c|c}
      %Condition                & H3.4R1.4 & H3.3R2.4 & H3.3R4.1 & H3.2R6.2 \\
			Condition                & A & B & C & D \\
		\hline	
			$p_0$ [MPa]              & 7.1   & 12.8 & 20   &  29 \\
			$T_0$ [K]               & 2840  & 2940 & 2790 &  2680\\
			$h_0$ [MJ$\cdot$kg$^{-1}$] &  3.4  & 3.3  & 3.3  &  3.2\\
			$M_{\infty}$ [-]        & 7.3   & 7.3  & 7.4  & 7.4\\  
			$T_{\infty}$ [K]        &  286  & 299  & 277  &  261\\
			$\rho_{\infty}$ [g$\cdot$m$^{-3}$] &  10   &  17  & 29   &   43\\
			$u_{\infty}$ [m$\cdot$s$^{-1}$]      & 2480  & 2530 & 2450 &  2400\\
      $Re_m$ [m$^{-1}$]   & $\rm{1.4 \cdot 10^6}$ &  $\rm{2.4 \cdot 10^6}$ & $\rm{4.1 \cdot 10^6}$ & $\rm{6.2 \cdot 10^6}$\\		
		\end{tabular}
	\caption{Averaged operating conditions of HEG at \unit{M=7.4}{} used in the presented study in combination with the wedge probe at an angle of attack of \unit{0}{\degree}.}
	\label{tab:HEGconditions}
\end{table}

\subsubsection{The Ludwieg Tube Facility at DLR}

\begin{figure}[htbp]
	\centering
 \includegraphics[width=1.00\textwidth]{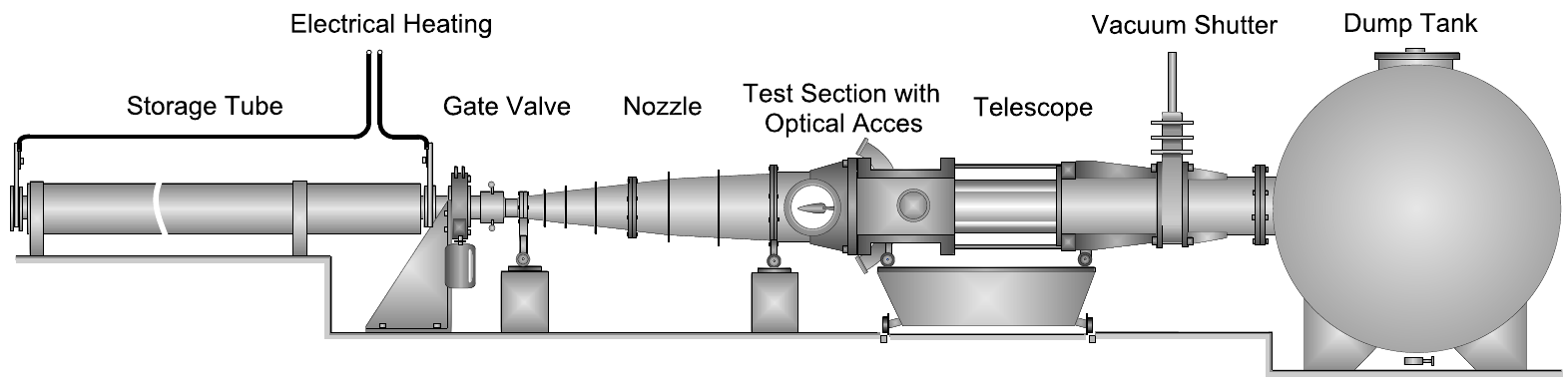}\\
 \caption{Schematic view of the Ludwieg tube facility DNW-RWG at DLR G\"ottingen.}
 \label{f:RWG}
\end{figure}

The Ludwieg tube facility DNW-RWG at DLR G\"ottingen, shown in figure \ref{f:RWG}, covers a Mach number range of 2 $\leq M_{\infty} \leq$ 7 and a unit Reynolds number range of 2 $\cdot$ 10$^{6}$ m$^{-1}$ $\leq Re_{m} \leq$ 11 $\cdot$ 10$^{7}$ m$^{-1}$. 
The facility uses an expansion tube as a high pressure reservoir, which is closed at one end and has a gate valve attached to the other end. The valve is followed by a supersonic nozzle, a test section and a dump tank. After opening the gate valve, the air flow is started by expansion waves traveling towards the closed end of the tube, where they are reflected. As long as these waves do not reach the nozzle throat, the test gas expands through the nozzle and the test section into the dump tank at nearly constant stagnation conditions. The Ludwieg Tube DNW-RWG has two tubes, an unheated tube A and a heated tube B, with a length of \unit{80}{m} each, resulting in a test time of about \unit{300-350}{ms}. The low operation costs, a relatively large test section, and the good optical access make this facility best suited for optical methods and heat flux measurements. \\

\begin{table}
\begin{minipage}[c]{0.49\textwidth}
\begin{tabular}{l|c|c|c}
Condition            &            Low       &            Medium              &            High       \\
\hline                                                                                                  
$p_0$ [MPa]     &            0.10             &            0.16          &            0.29          \\
$T_0$ [K]            &            259         &            259         &            258         \\
$M_\infty$ [-]  &            2.92       &            2.94       &            2.97       \\
$p_\infty$ [Pa] &            3071      &            4767      &            8259      \\
$T_\infty$ [K]   &            96           &            95           &            93           \\
$\rho_\infty$ [g$\cdot$m$^{-3}$]          &            112         &            175         &            308         \\
$u_\infty$ [m$\cdot$s$^{-1}$] &            573         &            574         &            575         \\
$Re_{m}$ [m$^{-1}$]         &            $9.0\cdot10^6$                &            $15.2\cdot10^6$             &            $27.4\cdot10^6$                \\
\end{tabular}
\caption{Applied test condition range, \textbf{Mach 3}, wedge probe \unit{AoA=0}{\degree}.}
\label{tab:RWGconditionsM3}
\end{minipage} \hfill
\begin{minipage}[c]{0.49\textwidth}
\begin{tabular}{l|c|c|c}
Condition            &            Low       &            Medium              &            High       \\
\hline                                                                                                  
$p_0$ [MPa]     &            0.42       &            1.70       &            2.89       \\
$T_0$ [K]            &            546         &            546         &            535         \\
$M_\infty$ [-]  &            5.98       &            5.98       &            5.98       \\
$p_\infty$ [Pa] &            272         &            795         &            1868      \\
$T_\infty$ [K]   &            67           &            67           &            66           \\
$\rho_\infty$ [g$\cdot$m$^{-3}$]          &            14           &            41           &            99           \\
$u_\infty$ [m$\cdot$s$^{-1}$] &            981         &            981         &            971         \\
$Re_{m}$ [m$^{-1}$]    &            $3.0\cdot10^6$                &            12.3$\cdot10^6$             &            21.5$\cdot10^6$             \\                                                                              
\end{tabular}
\caption{Applied test condition range, \textbf{Mach 6}, wedge probe \unit{AoA=10}{\degree}.}
\label{tab:RWGconditionsM6}
\end{minipage}
\end{table}

In the present experiments, tests at Mach 3 and 6 were conducted in the test condition range outlined in table \ref{tab:RWGconditionsM3} and \ref{tab:RWGconditionsM6}. For each Mach number the unit Reynolds number was varied by changing the reservoir pressure at approximately constant reservoir temperature. \\
Previous experimental studies to quantify the free-stream disturbances in the DNW-RWG Ludwieg tube at Mach 5 were conducted by means of hot wire and pitot probe measurements by Wendt \cite{Wendt1995}. A broadband (\unit{1-200}{kHz}) mass flow fluctuation of \unit{1.5}{\%} and a broadband \break (\unit{1-100}{kHz}) pitot pressure fluctuation of \unit{1.8}{\%} were reported. 

In the present study the circular Mach 6 and the 2-D Mach 3 nozzle were used, providing a nozzle exit diameter of \unit{0.5}{m} and a cross section area of \unit{0.5 \times 0.5}{m}, respectively. The free-stream probe was positioned on the nozzle axis in the nozzle exit plane.

\subsubsection{The Ludwieg Tube Facility at University of Braunschweig}

\begin{figure}[htbp]
	\centering
		\includegraphics[width=1.00\textwidth]{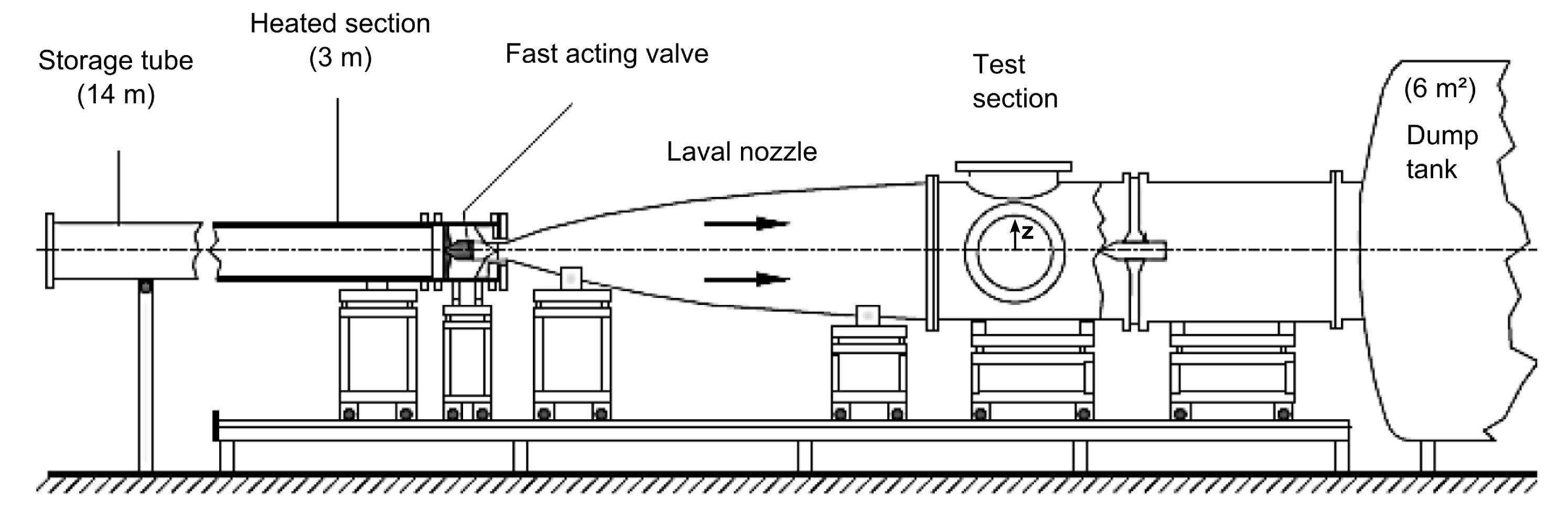}
	\caption{Schematic view of the Ludwieg tube HLB at TU Braunschweig.}
	\label{fig:HLB_wind_tunnel}
\end{figure}

\begin{figure}[htbp]
	\centering
		\includegraphics[width=0.50\textwidth]{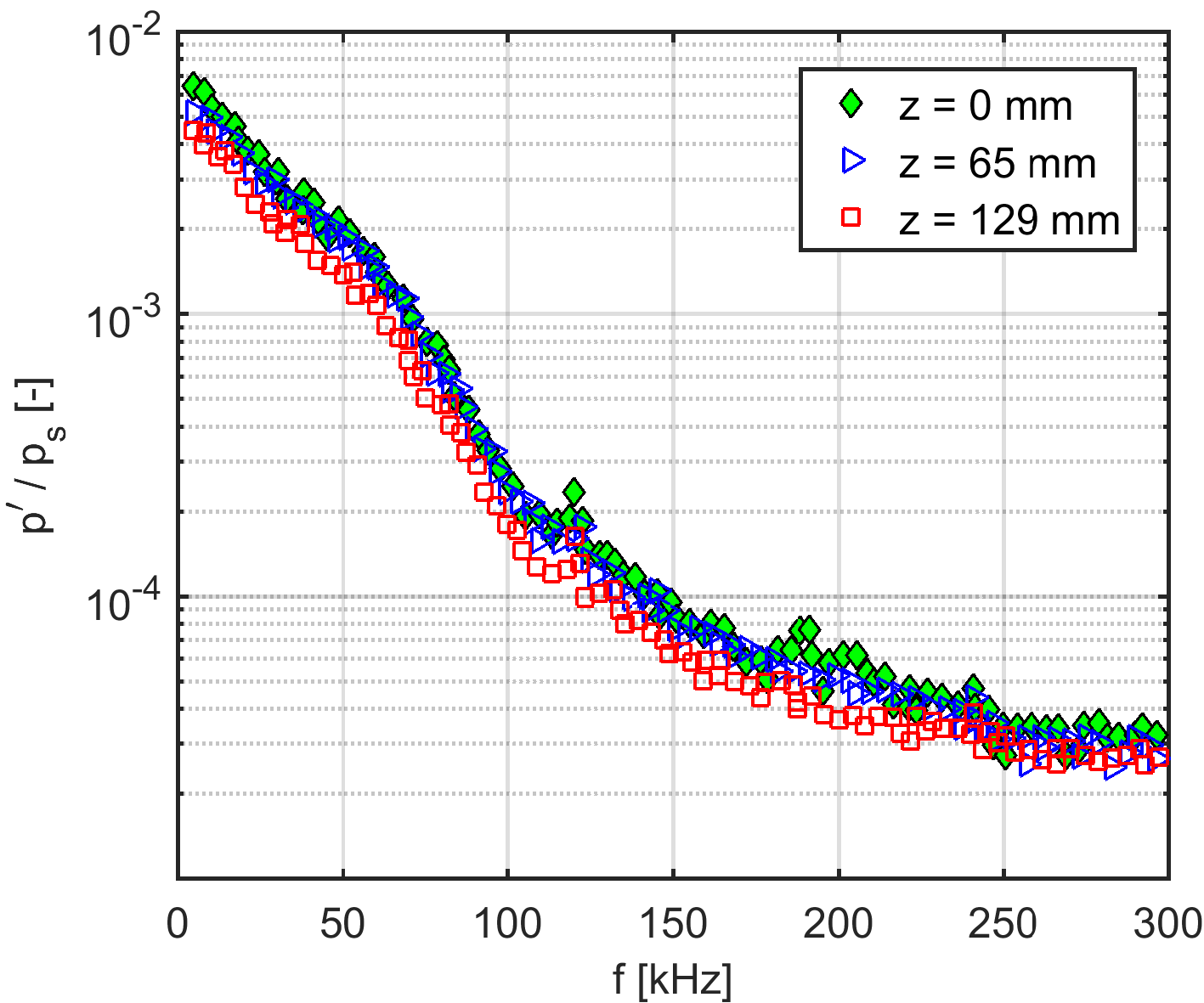}
	\caption{Spectra of normalized pitot pressure fluctuations measured at various off-axis positions in the center of the test section at an unit Reynolds number of $\unit{Re_m \approx 6 \cdot 10^6}{m^{-1}}$. Reconstructed from Heitmann et al. \cite{Heitmann2008}.}
	\label{fig:PitotPresFluc_HeitmannHLB}
\end{figure}

The hypersonic wind tunnel at the University of Technology Braunschweig (HLB) is a heated Ludwieg tube. It is divided into a high and a low pressure section. The high pressure section consists of a storage tube which can be pressurized in the range \unit{4-30}{bar}. To prevent condensation during the flow expansion the storage tube is partially heated. The low-pressure section consists of the Laval nozzle, the test section, the supersonic diffusor and a vacuum tank. The tunnel flow is initiated by opening a pneumatic fast-acting valve, which is located upstream of the nozzle throat. The Mach number at the nozzle exit is \unit{M_{\infty}=5.9}{}. The length of the storage tube limits the available test time to about \unit{80}{ms}. A schematic drawing of the facility is provided in figure \ref{fig:HLB_wind_tunnel}.
The temperature inside the storage tube is recorded at two positions close to the valve, one on the upper side and another on the lower side of the tube. The total temperature in the storage tube is obtained by averaging both signals. Previous measurements by Heitmann et al. \cite{Heitmann2008} assess the free-stream disturbance levels by means of pitot pressures probes. The tests revealed fluctuation levels of 1 and 3.6 \%, depending on the initial pressure and the position in the test section as depicted in figure \ref{fig:PitotPresFluc_HeitmannHLB}.

In the present studies a circular Mach 6 nozzle was used, providing a nozzle exit diameter of \unit{0.5}{m}. The test condition range is outlined in table \ref{tab:HLBconditions}. Unit Reynolds number variations were realized by changing the reservoir pressure at approximately constant total temperature. The free-stream probe was positioned \unit{85}{mm} above the nozzle centerline about \unit{300}{mm} downstream the nozzle exit plane. The position was chosen to allow a comparison with previously conducted cone probe measurements at this position conducted by Ali et al.\cite{Ali2014} and to avoid vorticity waves known to emanate from the plug valve upstream of the nozzle leading to significantly increase of the pressure fluctuations over the complete frequency range.\cite{Schilden2016} \\ 

\begin{table}[htbp]
\centering
\begin{tabular}{l|c|c|c}
Condition            &            Low       &            Medium              &            High       \\
\hline                                                                                                  
$p_0$ [MPa]     &            0.44     &            0.87     &            1.36     \\
$T_0$ [K]            &            448         &            464         &            463         \\
$M_\infty [-]$  &            5.9          &            5.9          &            5.9          \\
$p_\infty$ [Pa] &            306         &            612         &            956         \\
$T_\infty$ [K]   &            56           &            58           &            58           \\
$\rho_\infty$ [g$\cdot$m$^{-3}$]          &            19           &            36           &            57           \\
$u_\infty$ [m$\cdot$s$^{-1}$] &            887         &            903         &            902         \\
$Re_{m}$ [m$^{-1}$]         &            $4.3\cdot10^6$                &            $8.2\cdot10^6$                &            $12.8\cdot10^6$                \\                                           
\end{tabular}
  \caption{Applied test condition range in HLB in combination with a wedge probe an angle of attack of \unit{10}{\degree}.}
	\label{tab:HLBconditions}
\end{table}

\subsection{Wedge Probe Geometry and Instrumentation}
\label{sec:wedgeprobe}

\begin{figure}[hbtp]
	\centering
				\subfloat[Insert 1 at Mach 6, \unit{AoA=10}{\degree}]{\includegraphics[width=0.48\columnwidth]{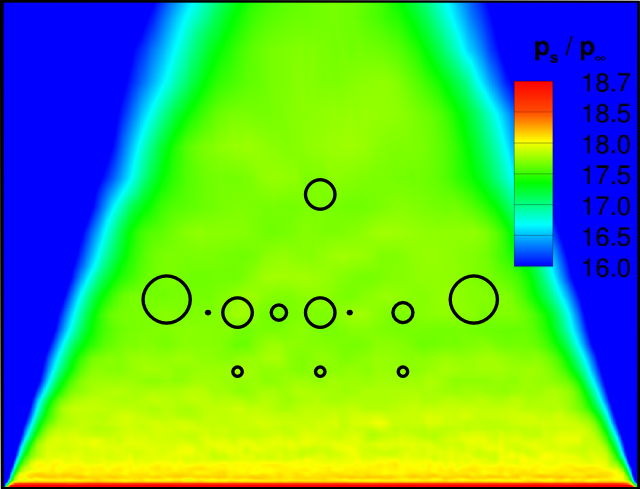}\label{fig:PCBInsertCFD}} \hspace{0.2cm}
				\subfloat[Insert 2 at Mach 3, \unit{AoA=0}{\degree}]{\includegraphics[width=0.48\columnwidth]{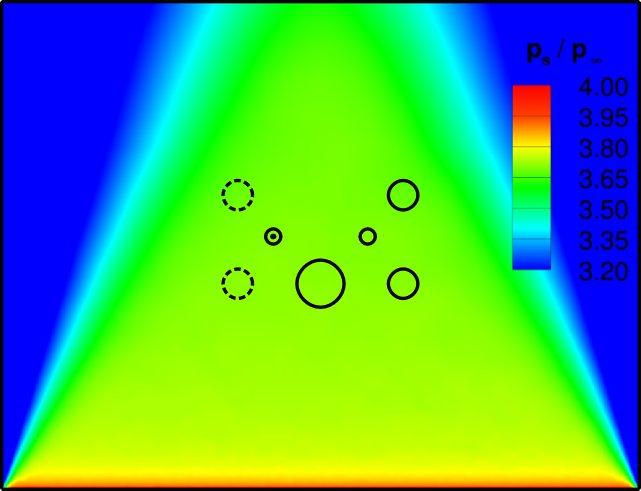}\label{fig:ALTPInsertCFD}} 
				\caption{Surface pressure normalized by the free stream static pressure on the instrumented wedge probe surface, \ref{fig:PCBInsertCFD}:  Mach 6 RWG at \unit{AoA=10}{\degree} and \ref{fig:ALTPInsertCFD}: Mach 3 in RWG at \unit{AoA=0}{\degree} . }
				\label{fig:InsertsCFD}
\end{figure}

\begin{figure}[hbtp]
	\centering
				\subfloat[Insert 1]{\includegraphics[width=0.45\columnwidth]{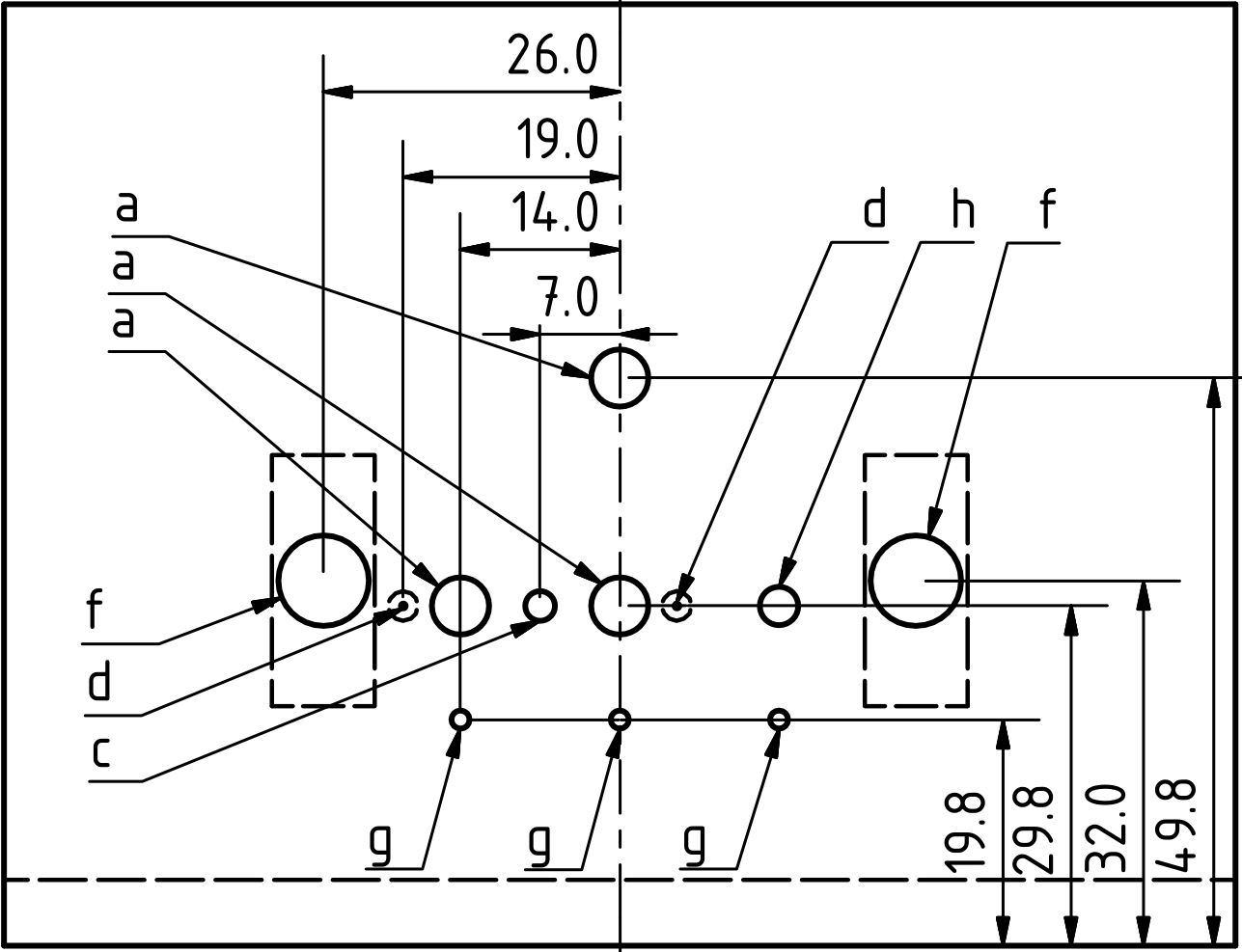}\label{PCBInsert}} \hspace{0.1cm}
				\subfloat[Insert 2]{\includegraphics[width=0.45\columnwidth]{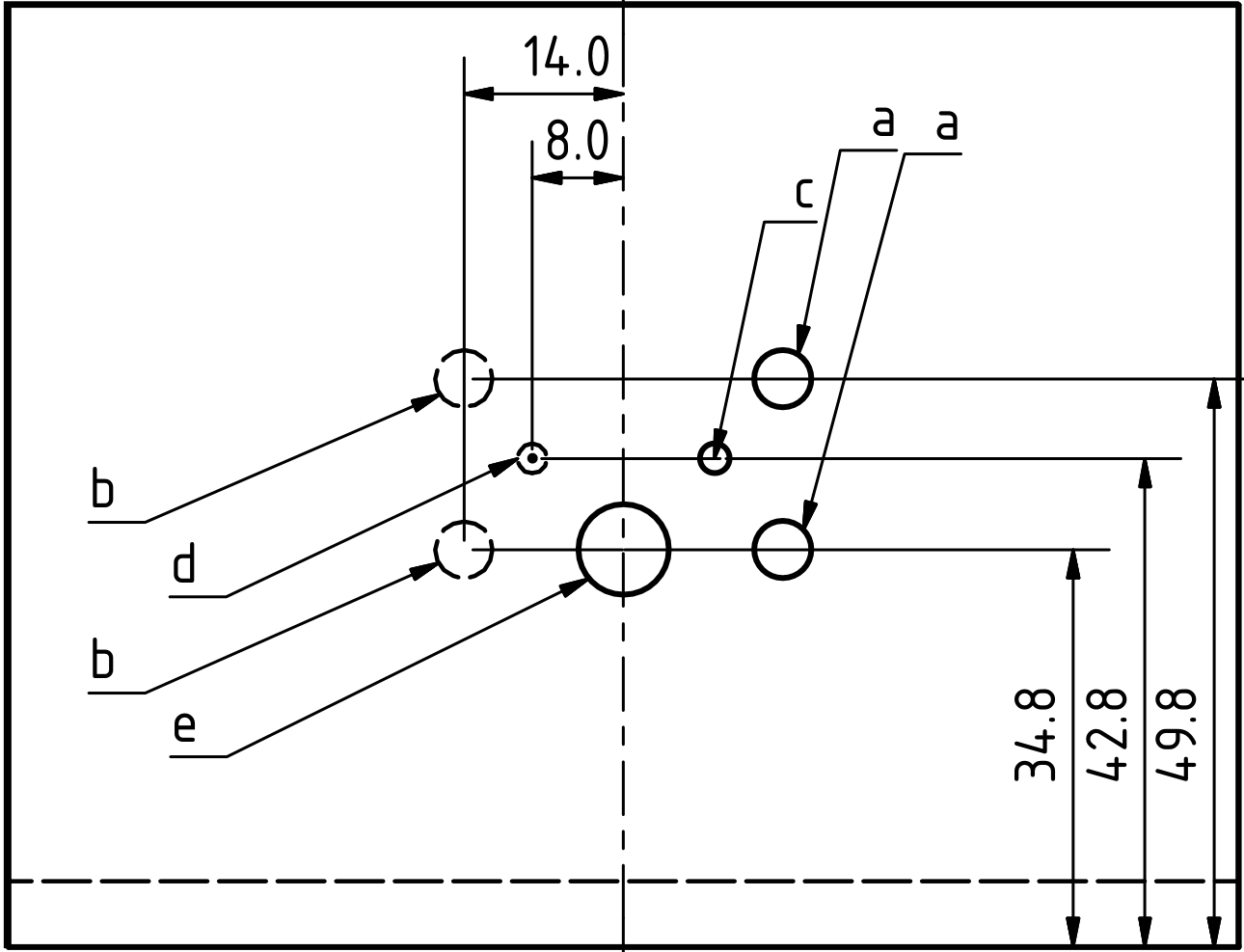}\label{ALTPInsert}} \hspace{0.1cm}
				\subfloat[Insert 3]{\includegraphics[width=0.45\columnwidth]{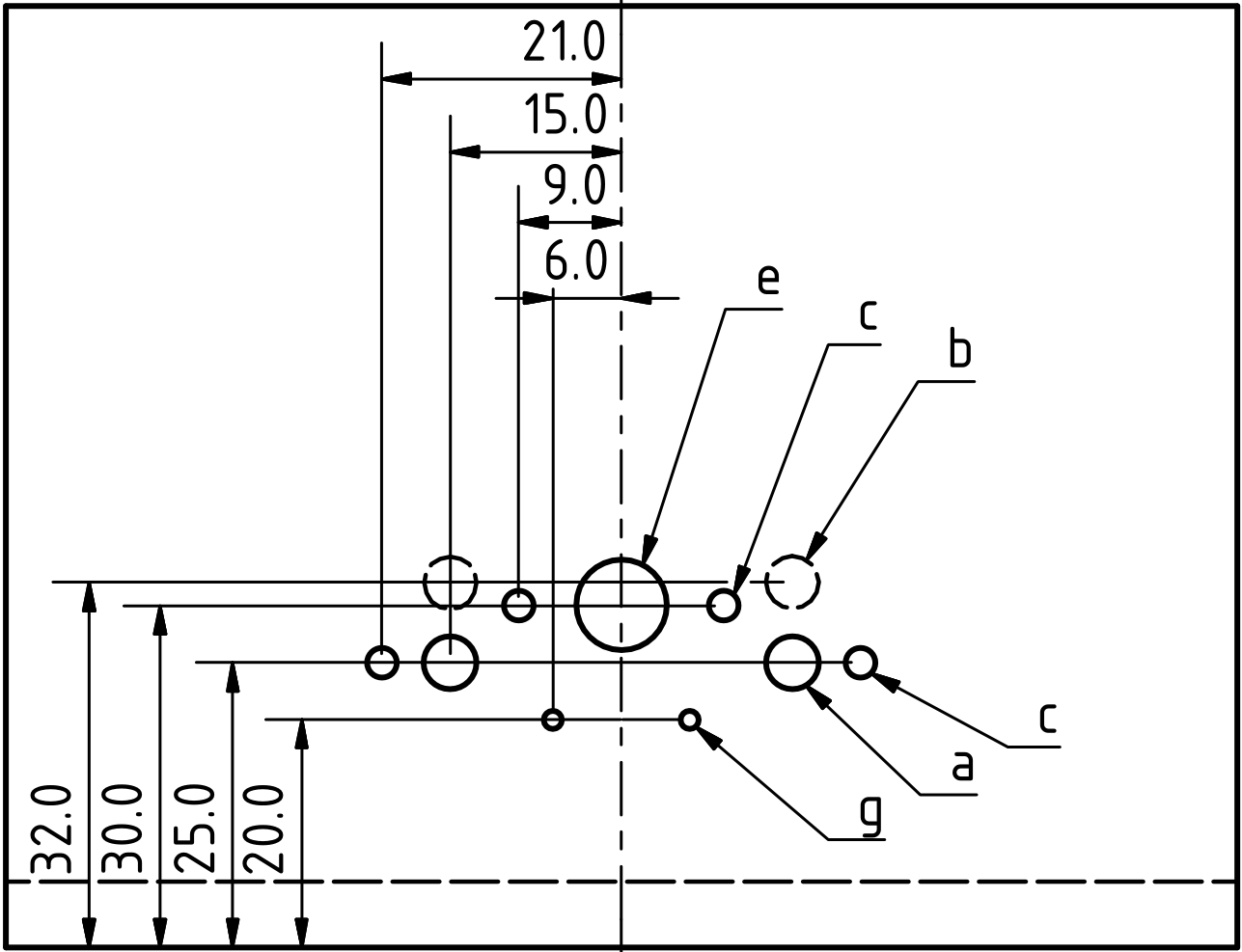}\label{HLBInsert}} 
				\caption{Wedge probe inserts. All dimensions are provided in millimeters. a - PCB pressure transducer, b - PCB pressure transducer without connection to the flow, c - flush mounted KULITE pressure transducers, d - KULITE pressure transducers behind cavity, e - ALTP heat flux transducer, f - low cost pressure transducer, g - type E coaxial thermocouples, h - thin film transducers}
				\label{fig:inserts}
\end{figure}

The main purpose of the study is to provide an easy-to-implement technique, to assess free-stream disturbances, suitable for harsh test environments as found, e.g., in shock tunnels. Therefore, the probe was designed to measure pressure, temperature and heat flux fluctuations at the surface of a slender body behind an oblique shock. The above requirements further imply that protective cavities around the transducers need to be minimized to ensure an undisturbed frequency response of the transducers. Regarding the probe dimensions a compromise was found, providing enough internal volume to integrate various types of transducers as close as possible to the leading edge while reducing the probe size to allow the integration into test sections in addition to a standard wind tunnel model. 
\begin{figure}[hbtp]
		\centering
			\includegraphics[width=0.95\textwidth]{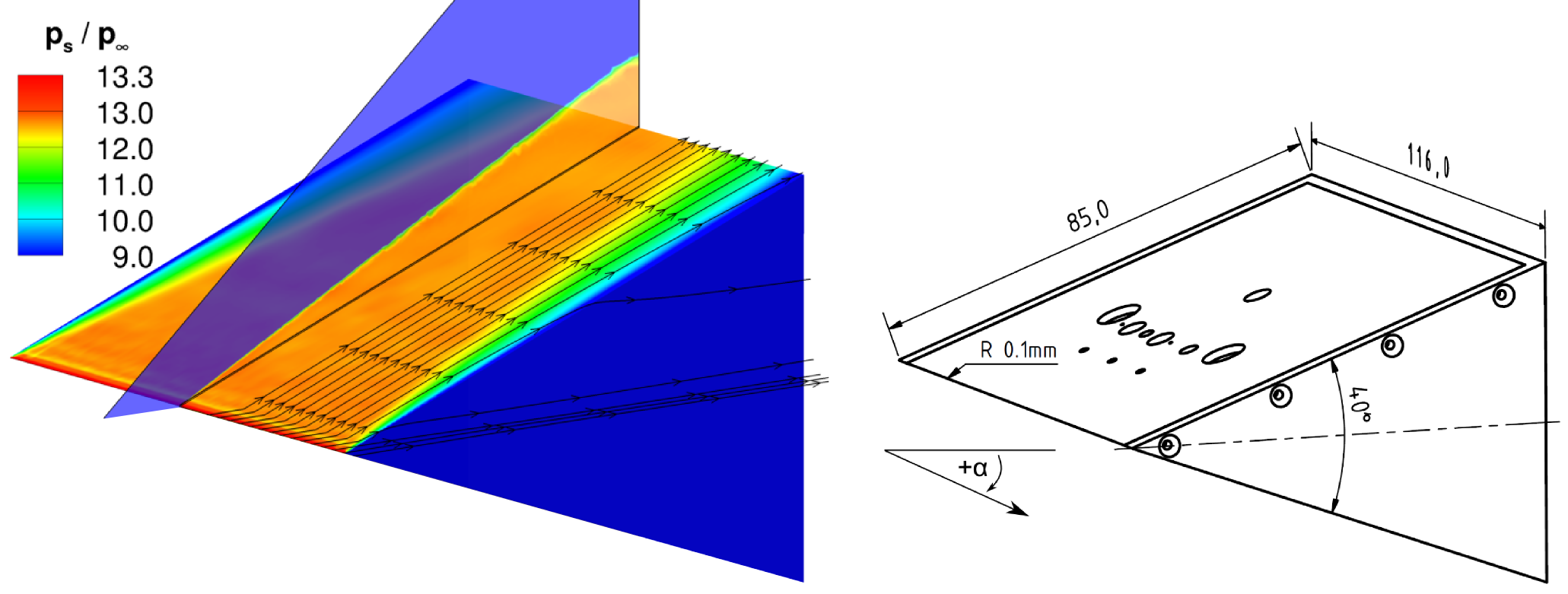}
		\caption{Surface pressure normalized by the free-stream static pressure on the wedge probe at \unit{AoA=0}{\degree} and Mach 7.4 in HEG, left. Basic probe dimensions, right. All dimensions are provided in millimeter.}
		\label{fig:3D_wedge_drawing}
	\end{figure}
The basic probe dimensions are provided in figure \ref{fig:3D_wedge_drawing} on the right. The probe was equipped with an exchangeable plane insert allowing the aerodynamically smooth integration of a wide range of transducers, while allowing the instrumentation to be adopted to different test conditions by changing the instrumented insert. Furthermore, the insert includes the leading edge of the probe which helps to avoid steps or gaps on the probe surface. The leading edge radius was chosen to be $\unit{0.1}{mm}$, allowing repeatable manufacturing. The probe can be used at different angles of attack to increase the signal-to-noise ratio in low pressure or low temperature environments. Since the probe extension is limited in the spanwise direction, side effects, dependent on the angle of attack and the Mach number, need to be considered. To assess the effect of the limited probe extension 3D RANS computations at Mach 3, 6 and 7.4 were conducted using the DLR TAU code.\cite{Mack2002,Gerhold2005,Schwamborn2006} Figure \ref{fig:3D_wedge_drawing} (left) depicts selected stream lines starting just above the leading edge of the probe and the normalized surface pressure distribution, $p_s/p_{\infty}$, on the probe at zero degree angle of attack and Mach 7.4 in HEG. The computations reveal that, although side effects are present at the probe limits, an undisturbed region of constant surface pressure exists in which the instrumentation is placed. Additional computations were conducted covering the lower Mach number test conditions applied in RWG. Figure \ref{fig:PCBInsertCFD} depicts the normalized surface pressure distribution on the probe at Mach 6 and \unit{10}{\degree} angle of attack. Figure \ref{fig:ALTPInsertCFD} provides the corresponding results for Mach 3 and zero degree angle of attack. Both computations prove that the instrumentation is located well within the region of undisturbed flow.\\
Figure \ref{fig:inserts} depicts the three inserts used in the present study. The first two inserts hold a range of different transducers such as standard and high frequency pressure transducers, coaxial thermocouples, thin film gages and ALTP heat flux transducers. While insert 1 and 2 were used in the initial tests in HEG and RWG to assess the transducer properties and limitations, insert 3 was designed based on the experience gathered in the preceding tests. \\
The tests revealed that the probe design allows the pressure transducers to be installed without a protective cavity. Even in harsh test environments, as present in HEG, no transducers were lost due to particle impact or overheating. Since cavities were shown to alter the frequency spectra by damping the high frequency content, the final insert uses flush-mounted pressure transducers only. Owing to the design of the low cost pressure transducers an installation without a cavity is not possible. As a consequence, these transducers were not used in the most recent probe layout and are not discussed in the present paper. \\
\indent Furthermore, tests in the RWG Ludwieg tube revealed that the frequency response of thin film gages is not suited for high frequency measurements in cold hypersonic tunnels. Thus, thin film gauge results are excluded from the discussion in the present paper. The same holds for measurements using coaxial thermocouples.

\section{Power Spectrum and RMS Estimation}
\label{sec:PSD2RMS}

Disturbance measurements in different wind tunnels are often compared by means of the overall RMS of for instance the pitot pressure. This approach provides the advantage of being easy to apply and being independent of the signal length and the sampling frequency assuming the signal RMS is invariant and sufficiently oversampled with respect to the Nyquist criterion to allow amplitude measurements. However, the approach has the disadvantage that natural transducer resonances are not considered and that no information on the frequency dependence of the disturbance amplitudes is provided. Since the hypersonic transition process largely depends on the high frequency content of a disturbance environment it is appropriate to provide the frequency spectra of a disturbance environment rather than a single RMS value, which is dominated by the low frequency content of the signal. 
In the present study the power spectrum and its relation to the RMS, provided by Press et al. \cite{Press1992}, is used to quantify the free-stream disturbances measured in the three wind tunnels as function of frequency. \\ 

The power spectrum (PS) can be estimated using the discrete Fourier transformation. Supposing a time signal $x(t)$ sampled at $N$ points at a constant sampling interval $\Delta t$ values in the range $x_0 ... x_{N-1}$ are produced. The time range of the signal is $T$ with $T = (N-1) \Delta t$ and the sampling frequency is $f_s$ leading to the frequency resolution of $\Delta f = f_s / N$. The discrete Fourier transform of $x$ is defined as
\begin{equation}
	X_k=\overset{N-1}{\underset{j=0}{\sum}} x_j \,e^{-2 \pi i j k / N} \;, \text{for}~k=0,...,N-1.  \nonumber
\end{equation}
The periodogram based estimate of the power spectrum at $N/2+1$ frequencies is defined (Press et al. \cite{Press1992}) as 
\begin{align}
 &PS(0)=PS(f_0)= \frac{1}{N^2} \left| X_0 \right|^2  \\
 &PS(f_k) = \frac{2}{N^2} \left| X_k \right|^2 , \text{for}~k=1,2,...,\frac{N}{2}-1 \label{equ:PSestimation} \\
 &PS(f_{Ny}) = PS(f_{N/2}) = \frac{1}{N^2} \left| X_{N/2} \right|^2.
\label{equ:PSestimate}
\end{align}
The first element of the power spectrum, $PS(f_0)$, corresponds to zero frequency and thus is the average of the time series. Since the mean value of the signal is usually subtracted before computing the Fourier transform this term can be neglected. The last element, $X_{N/2}$, corresponds to the Nyquist frequency, $f_{Ny}=f_{N/2} = f_s/2$, and requires special treatment as well. However, in practice it is removed by anti-aliasing filters before the A/D signal conversion and thus can be neglected. Hence, the power spectrum can be estimated following equation \ref{equ:PSestimation} for $k=1,...,\frac{N}{2}$. Furthermore, Parseval$^\prime$s theorem states that the total energy of a signal $x(t)$ in the time domain equals the total energy of its Fourier transform $X(f)$ in the frequency domain. The following form of Parseval$^\prime$s theorem holds for discretized signals (Smith\cite{Smith1997}),
\begin{equation}
	\overset{N-1}{\underset{i=0}{\sum}}|x_{i}|^2 =\frac{2}{N}\overset{\frac{N}{2}}{\underset{k=0}{\sum}}|X_k|^2.
\end{equation}
The theorem implies that the root mean square (RMS) can be formulated as
\begin{align}
	RMS(x)=&\sqrt{\frac{2}{N^2}\overset{\frac{N}{2}}{\underset{k=0}{\sum}}|X_k|^2}\, 
\overset{Eq. \ref{equ:PSestimation}}{=} \sqrt{\overset{\frac{N}{2}}{\underset{k=0}{\sum}}\text{PS}(f_k)}. 
\label{equ:FFT2RMS}
\end{align}
Consequently, the root mean square resulting from a single bin of the width $\Delta f$ centered around $f_k$ in the spectrum is
\begin{align}
	RMS(x,[f_k \mp {\Delta f}/{2}])&=\sqrt{\text{PS}(f_k)} = \frac{\sqrt{2}}{N}|X(f_k)|. 
\label{equ:rms1def}
\end{align}
Finally, the RMS of a time signal in the frequency range $f_m ... f_n$ is derived as 
\begin{align}
	RMS([f_m, f_n])&=\sqrt{\frac{2}{N^2}\overset{k(f_n)}{\underset{k(f_m)}{\sum}}|X(f_k)|^2} \\ 
	&=\sqrt{\overset{k(f_n)}{\underset{k(f_m)}{\sum}}PS(f_k)},
	\label{equ:PStoRMS}
\end{align}
with $m<n$ and $m,n=1,...,\frac{N}{2}$. Equation \ref{equ:rms1def} is also known as the linear or amplitude spectrum (AS). It is used in section \ref{sec:expresults} to provide the signal RMS based on a $\unit{1}{kHz}$ frequency range by summarizing its entries according to equation \ref{equ:PStoRMS}. The summation over a frequency range of $\unit{1}{kHz}$ provides the advantage of naturally smoothing the spectral distributions obtained in short duration facilities, while respecting the frequency resolution relevant for short test time hypersonic facilities. Furthermore, the approach allows the extraction of RMS information from a signal in a specific frequency range of interest, e.g. for hypersonic transition studies.

\section{Numerical method} \label{Chapter:Numerical method}

\subsection{Governing equations} \label{Section:Governing equations}

%\begin{itemize}
%\item	\red{Are the DNS 2D, because the wedge is (largely) a two-dimensional configuration? How about the circular wind tunnel wall emitting noise in all directions - how do you account for the different directions in the side-ways direction?}
%\end{itemize}

Numerical simulations of the Navier-Stokes equations for compressible flows, under the assumption of a perfect gas, were carried out in a two-dimensional (2D) reference system for the cylinder-wedge geometry of the measurement probe. The simplifying 2D assumption is justified in this case, apart from the 2D geometry of the measurement probe and the negligible side effects (as seen in Section \ref{sec:PSD2RMS}), since a) we are inserting small amplitude free-stream disturbances (to study the linear regime), which prevent the formation of nonlinearities, which might enhance the rapid generation and growth of 3D instability modes, and b) we are analyzing the early nose region, namely the region upstream the second mode neutral point. In the presence of three-dimensional (3D) free-stream acoustic waves (Cerminara \cite{Cerminara2017}), the response was found to be dominated by 2D modes. \\   

The set of non-dimensional conservation equations written in 2D curvilinear coordinates is 
% We consider numerical solutions of the three-dimensional Navier-Stokes equations for compressible flows, written in conservation form, under the assumption of perfect gas. 
	\begin{equation}
	\frac{{\partial}J\textbf{Q}_c}{{\partial}{t}}\ +  \frac{{\partial}\textbf{F}}{{\partial}{\xi}}\ + \frac{{\partial}\textbf{G}}{{\partial}{\eta}}\ = \ 0,
	\end{equation}
where ($\xi$, $\eta$) are the curvilinear coordinates, while the Cartesian coordinates are $x=x(\xi,\eta)$ and $y=y(\xi,\eta)$ and the Jacobian is given by $J= \det||{\partial}{(x,y)}/{\partial}{(\xi,\eta)}||$. 
In the equation above, $\textbf{Q}_c=\left[\rho \ \rho u \ \rho v \ \rho E \right]^T$ is the vector of the conservative variables, while $\textbf{F}$ and $\textbf{G}$ are the vectors of the fluxes. 

The terms $\rho$ , $\rho u$ , $\rho v$ , and $\rho E$ are the non-dimensional conservative variables, where $\rho$ is the density, $u$ and $v$ are the velocity components respectively in the $x$, and $y$ directions, and $E$ is the total energy per unit mass. 

The symbol $*$ is used in the present section to denote dimensional values. Velocity components are normalized with the free-stream main velocity ($U_\infty^{*}$), density with the free-stream density ($\rho_\infty^{*}$), viscosity with the free-stream dynamic viscosity ($\mu_\infty^{*}$), temperature with the free-stream temperature ($T_\infty^{*}$), total energy with the square of the free-stream mean velocity ($U_\infty^{*2}$), while the pressure and viscous stresses are normalized with $\rho_\infty^{*} U_\infty^{*2}$. The dimensional nose radius ($R^{*}$) is chosen to normalize length scales, while the time scales are normalized with respect to a characteristic time ($R^{*}/U_\infty^{*}$), based on the velocity of the undisturbed flow and on the characteristic length. The relevant dimensionless quantities are \textit{Re}, \textit{Pr}, \textit{M}, and $\gamma$, which are respectively the Reynolds, Prandtl and Mach numbers, and the ratio of specific heats ($\gamma = c_p^{*} / c_v^{*}$). The Reynolds number is defined with respect to the nose radius, as $Re = (\rho_\infty^{*} U_\infty^{*} R^{*})/ \mu_\infty^{*} $; the Prandtl number is set to 0.72 for air, and $\gamma$ is equal to 1.4, as we are considering a calorically perfect gas model. The dynamic viscosity is, in turn, expressed in terms of temperature by Sutherland's law 
	\begin{equation}
	\mu =  T^{3/2} \frac{1 + C}{T + C} \ , 
	\end{equation}
where the constant C represents the ratio between the Sutherland's constant (set to 110.4 K) and a reference temperature ($T_\infty^{*}$). 

During the computations the inlet boundary condition is either a fixed inflow condition (in the steady state simulations) or has a prescribed time-dependent form according to an acoustic wave function detailed in the following section. On the body surface an isothermal wall boundary condition is used, with wall temperature fixed at a constant value dependent on the particular case. This is appropriate in modeling experiments in short-duration hypersonic wind tunnels, where the wall temperature is subject to small changes only.

\subsection{Modeling of planar acoustic waves} \label{Section:Modeling of planar acoustic waves}

\begin{figure}[!h]% order of placement preference: here, top, bottom
 \centering
 \includegraphics[width=0.5\textwidth]{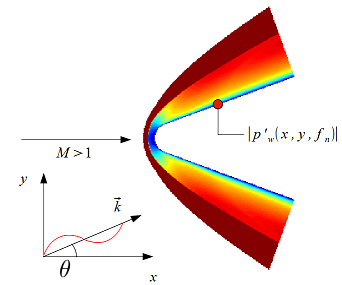}
 \caption{Sketch of the planar acoustic waves and of the computational domain near the nose region. The $u$ velocity field is shown for illustration purposes.}
 \label{fig:acoustic_waves}
\end{figure}

Figure \ref{fig:acoustic_waves} shows a sketch of the planar acoustic waves traveling in the direction of the wave vector $\overrightarrow{k}$, with an inclination angle $\theta$ with respect to the positive $x$ axis of the Cartesian reference system. The wave vector ($\overrightarrow{k}$) indicates a general propagation direction of the acoustic waves in the $xy$-plane, and $|p'_{w}(x,y,f_{n})|$ denotes the absolute value of the pressure fluctuations on the wall at a generic ($x$,$y$) point for a generic frequency ($f_{n}$, with $n=1,2,..,N$) inside the range of considered frequencies.

The free-stream perturbation amplitudes of the velocity components ($|u'|$, $|v'|$), pressure ($|p'|$) and total energy ($|E'|$) are expressed in terms of the density perturbation amplitude ($|\rho'|$) by means of the following relations, derived from the linearized Euler equations under the assumption of small perturbations

	\begin{equation}
	|u'| = \frac{1}{M}|\rho'| \cos\theta \ ,\ |v'| = \frac{1}{M}|\rho'| \sin\theta \ , \ |p'| = \frac{1}{M}|\rho'|,
	\end{equation}
    
\noindent
and

	\begin{equation}
	|E'| = \frac{1}{M}|\rho'|\left(\frac{1}{\gamma M} +  \cos \alpha \cos\theta + \sin \alpha \sin\theta \right),
	\end{equation}
\\[0.2cm] where $\alpha$ denotes the angle of attack. The inclination angle of the acoustic waves ($\theta$) is considered positive for waves impinging from below. Hence, we can impose the free-stream perturbation amplitude for the density and make use of the relations above to fix the fluctuation amplitude of the physical quantities. The relations for the pressure fluctuation amplitude and the velocity component fluctuation amplitudes are consistent with the dispersion relations shown in the work of Egorov \textit{et al.} \cite{egorov2006}, while a derivation of the total energy perturbation is shown in Cerminara and Sandham \cite{Cerminara2015}. Once the amplitude is assigned, the free-stream perturbation of the density as a function of time and the Cartesian coordinates, for the case of multiple frequencies, is expressed as 
	\begin{equation}
	\rho'(x,y,t) = |\rho'| \sum_{n=1}^{N} \cos\left(k_{nx}x + k_{ny}y - \omega_{n}t + \phi_{n}\right), 
	\end{equation}
where $k_{nx}$ and $k_{ny}$ are the wavenumbers respectively in the $x$ and $y$ directions, $\omega_{n}$ is the angular frequency and $\phi_{n}$ is the phase angle of the acoustic wave for the $n^{th}$ frequency, while $N$ represents the total number of frequencies of the wave spectrum. These terms are, in turn, expressed by the following relations 
	\begin{align}
	k_{nx}&=|k_{n}|\cos\theta \ ; \ k_{ny}=|k_{n}| \sin\theta \ ; \\ 
	|k_{n}| &= \frac{\omega_{n}}{ \cos\theta \pm 1/M} \ ; \\
	\omega_{n} &= n\omega_{1} = 2\pi n f_{1} \ .
	\end{align}
Here, $|k_{n}|$ is the magnitude of the wave vector for the $n^{th}$ frequency, which depends on the angle $\theta$ since the convection velocity of the acoustic waves (as illustrated in figure \ref{fig:acoustic_waves}) is the projection of the mean free-stream velocity along the propagation direction of the acoustic waves. With $f_{1}$ we refer to the smallest frequency of the complete spectrum, and each imposed frequency is a multiple of $f_{1}$. The plus sign in the denominator of $|k_{n}|$ is applicable for fast acoustic waves, whereas the minus sign is for slow waves. The perturbations of the other variables are easily obtained from the density perturbation function and the relations for the amplitudes listed above. The vector of the conservative variables at the inflow boundary in the unsteady computations is given by
	\begin{equation}\label{eq:unsteady_quantities}
	\textbf{Q}_{c}^{U} = 
	\left[ {\begin{array}{c} 
		 {\rho}_{\infty} + \rho' \\
		 ({\rho}_{\infty} + \rho')({u}_{\infty} + u') \\
		 ({\rho}_{\infty} + \rho')({v}_{\infty} + v') \\
		 ({\rho}_{\infty} + \rho')({E}_{\infty} + E') 
	\end{array} } \right] \ ,
	\end{equation}
where the subscript $\infty$ denotes free-stream mean values of the physical quantities.

\subsection{Code features} \label{Section:Code features}

The DNS computations are carried out with the in-house SBLI (Shock-Boundary-Layer-Interaction) code, in which shock-capturing is applied as a filter step to the solution obtained through the base scheme at the end of each time integration cycle. The code uses fourth-order central finite difference scheme for space discretization and makes use of an entropy-splitting method (Yee \textit{et al.}\cite{yee2000}) to improve the nonlinear stability of the high-order central scheme. Near the wall a fourth order Carpenter boundary scheme (Carpenter \textit{et al.}\cite{Carpenter1999}) is chosen, while for time integration, a third order Runge-Kutta scheme is used. The shock-capturing scheme consists of a second-order TVD (total variation diminishing)-type algorithm, with a particular compression method (Yee \textit{et al.}\cite{yee1999}) in order to add the dissipation in an efficient way into the flowfield. The scheme is supplemented with the Ducros sensor (Ducros \textit{et al.}\cite{ducros1999}), which additionally limits the numerical dissipation in the boundary layer. The code has been set up to run in parallel using MPI libraries. More details, together with a validation of the code can be found in the work of De Tullio \textit{et al.}\cite{detullio2013}, where DNS results are compared with PSE (Parabolized Stability Equations) results for the case of transition induced by a discrete roughness element in a boundary layer at Mach 2.5.

\subsection{Flow conditions and settings of the numerical simulations}

Table \ref{tab:flow_conditions} shows the flow conditions of the simulated cases, namely the free-stream Mach number ($M$), unit Reynolds number ($Re_{m}$), stagnation temperature ($T_0^{*}$), free-stream temperature ($T_\infty^{*}$), free-stream pressure ($p_\infty^{*}$), wall temperature ratio ($T_w^{*}/T_\infty^{*}$), angle of attack ($AoA$), and angle of incidence of the acoustic waves ($\theta$). The flow conditions reproduce the free-stream of selected experimental tests carried out in the HEG and RWG facilities. 

\begin{table*}% no placement specified: defaults to here, top, bottom, page
 \begin{center}
  \begin{tabular}{c|c|c|c|c|c|c|c|c|c}
       Case & Facility & $M$ & $Re_{m}$ [m$^{-1}$] & $T_0^{*}$ [K] & $T_\infty^{*}$ [K] & $p_\infty^{*}$ [Pa] & $T_w^{*}/T_\infty^{*}$ & $AoA$ [\degree] & $\theta$ [$\degree$] \\\hline
      1 & HEG & 7.3 & 4.4$\times10^{6}$ & 2740 & 234.034 & 2004.301 & 1.273 & 0  &  0 \\
      2 & HEG & 7.3 & 1.4$\times10^{6}$ & 2680 & 228.909 & 619.337 & 1.302 & 0  &  0 \\
      3 & HEG & 7.3 & 1.4$\times10^{6}$ & 2680 & 228.909 & 619.337 & 1.302 & 0  & 10 \\
      4 & RWG & 6.0 & 6.3$\times10^{6}$ & 559  & 68.571 & 588.852 & 4.346 & 10  &  0 \\
      5 & RWG & 3.0 & 12.0$\times10^{6}$ & 258  & 92.538 & 3588.5 & 3.22 & 0  &  0 
  \end{tabular}
   \caption{Flow conditions of the numerical cases.}
  \label{tab:flow_conditions}
 \end{center}
\end{table*}

\begin{figure}[hbtp]
 \centering
 \includegraphics[width=0.6\columnwidth]{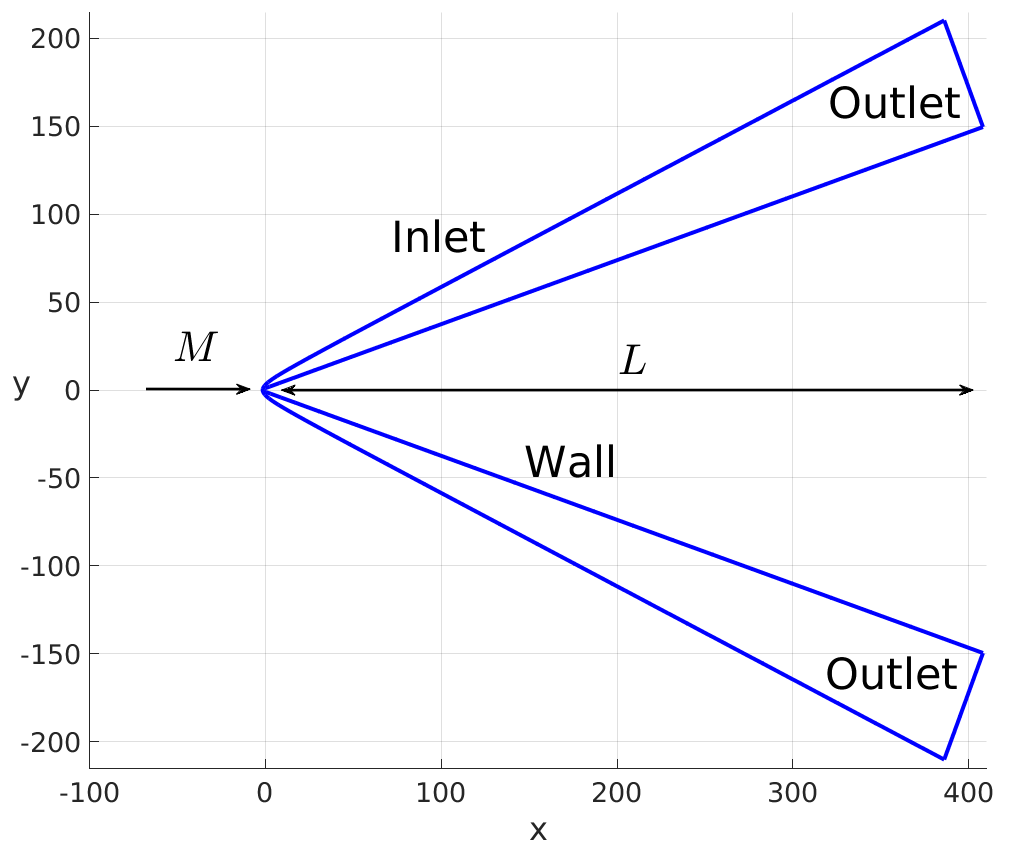}
 \caption{Limits of the computational domain. All dimensions were normalized by the probe nose radius of \unit{R^*=0.1}{mm}. }
 \label{fig:geometry}
\end{figure}

\noindent
A sketch of the geometry of the computational domain in the physical space is presented in figure \ref{fig:geometry}. The domain is adapted to both the body and the shock shape, based on the grid generation method by Bianchi \textit{et al.} \cite{bianchi2010}. The nose radius is \unit{R^*=0.1}{mm}, and the half-wedge angle is set to $20^{\circ}$, according to the geometrical details of the probe used in the experiments. 

The computational domain has a non-dimensional length ($L$) of 410 nose radii in the streamwise direction, in order to include the points along the wall where the transducers were located in the experimental tests. The grid size for cases 1 to 3 in \ref{tab:flow_conditions} is 2244$\times$150 (where 2244 is the number of points along the streamwise direction, \textit{i}, and 150 is the number of points in the wall-normal direction, \textit{j}), while the grid size for cases 4 and 5 is 2244$\times$200. A grid sensitivity study was shown in Cerminara \cite{Cerminara2017}.

Acoustic waves were chosen as free-stream disturbances, since they are known to represent the dominant disturbance type generated by turbulent boundary layers on the nozzle walls of hypersonic wind tunnels (Duan \textit{et al.}\cite{Duan2014}). For each case listed in Table \ref{tab:flow_conditions}, unsteady simulations were performed with both fast and slow acoustic waves. %In particular, after the computation of the baseflow, planar fast/slow acoustic waves were inserted into the domain through a time-periodic boundary condition at the inlet boundary (according to equation \ref{eq:unsteady_quantities}), and 
The unsteady simulations were performed until periodic convergence of the solution was reached. For each frequency in equation \ref{eq:unsteady_quantities}, an amplitude level of 10$^{-4}$ was chosen for the free-stream density fluctuation, in order to guarantee linear results throughout the domain.

In the present numerical formulation, the frequency is normalized with the nose radius ($R^{*}$) and the free-stream mean velocity ($U^{*}$), as $f=f^{*}R^{*}/U_{\infty}^{*}$, where $f^{*}$ is the dimensional frequency. For all the cases with fast acoustic waves, a set of $N=10$ frequencies ranging from 50 kHz to 500 kHz was imposed. For slow acoustic waves, the frequency range is case-dependent, so that the frequency resolution is improved in the lower frequency range for the HEG cases (Cases 1 to 3). Therefore, for slow waves 10 frequencies were inserted in the ranges 20 kHz to 200 kHz for Cases 1 and 4, 25 kHz to 250 kHz for cases 2 and 3, and 50 kHz to 500 kHz for Case 5.

\section{Experimental Results}
\label{sec:expresults} 

The wedge probe was used in a series of tests in the reflected shock tunnel HEG (at low enthalpies, $\unit{\approx 3}{MJ/kg}$), the DNW-RWG Ludwieg tube and the HLB Ludwieg tube. In the latter two the probe was used at an angle of attack of $\unit{10}{\degree}$ at Mach 6 to increase the signal to noise ratio. All transducers discussed in the present section provided repeatable results, allowing a determination of the spectral distribution of the surface pressure up to approximately $\unit{300}{kHz}$ depending on the facility and the test condition. \\

\begin{figure}[htbp]
	\centering
		\includegraphics[width=1.00\textwidth]{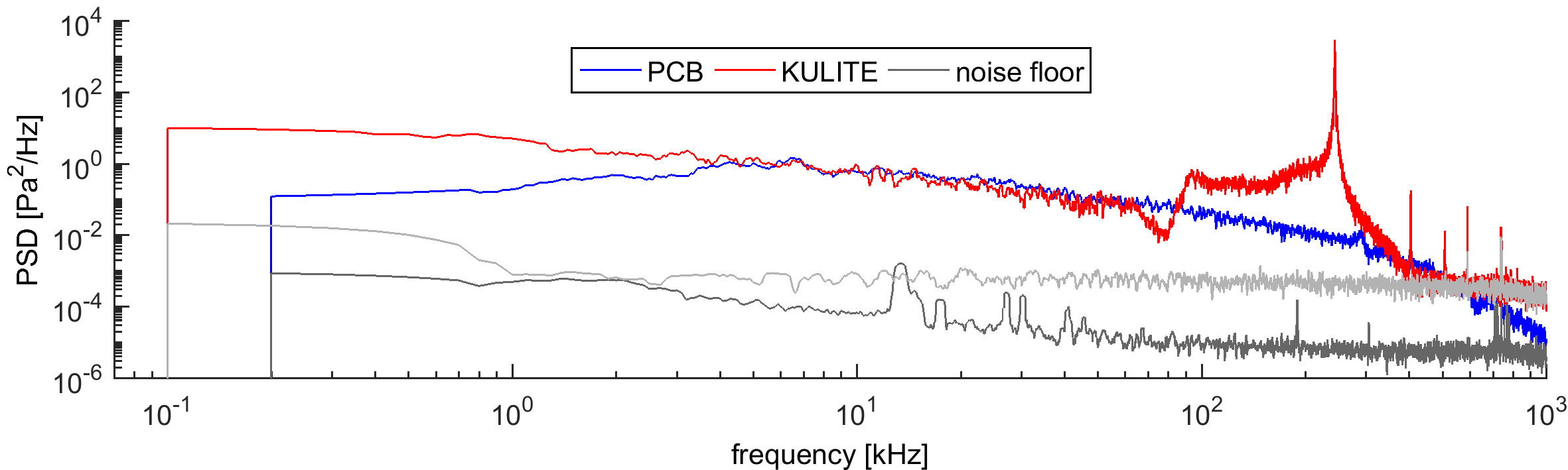} 
		\caption{Power spectral density of a piezoresistive pressure transducer (KULITE) and piezoelectric pressure transducer (PCB) indicating transducer resonance and the low frequency response, respectively.}
		\label{fig:PSD_exampleKU_PCB_1}
\end{figure}

A major goal of the present study is the quantitative evaluation of the disturbance environment over a wide frequency range. However, not every transducer can be used over the full frequency range. Transducer resonance frequencies or a low frequency response, as typically existing in piezoelectric transducers, alter the signal and its spectra as depicted in figure \ref{fig:PSD_exampleKU_PCB_1}. To overcome this drawback, different transducer types were combined in the present study. The approach takes advantage of the high precision of the piezoresistive transducers (e.g. from KULITE\textsuperscript{\textregistered}) at low frequencies and the high bandwidth of the piezoelectric pressure transducers of PCB\textsuperscript{\textregistered}. 
Possible uncertainties in the nominal piezoelectric transducer calibration could be compensated by applying an in-situ calibration against a calibrated piezoresistive transducer. In the example provided in figure \ref{fig:PSD_exampleKU_PCB_1} the power density spectra of both transducers overlap well above the low frequency limit of the PCB transducer and below the frequency at which the KULITE spectrum is altered (\unit{\approx 60}{kHz}) due to its resonance frequency (\unit{\approx 240}{kHz}). Thus, in the present case an adaptation 
% or the word "alteration" (instead of "adaptation") may be more appropriate?
of the PCB sensitivity is not necessary. In all subsequent evaluations the piezoresistive transducers were used to evaluate the low frequency range, in the present case below \unit{60}{kHz}. The piezoelectric transducers were used to evaluate the high frequency range between \unit{11}{kHz} and \unit{1000}{kHz}. \\

\begin{figure}[hbtp]
	\centering
			  \subfloat[HEG Mach 7.4]{\includegraphics[width=0.99\textwidth]{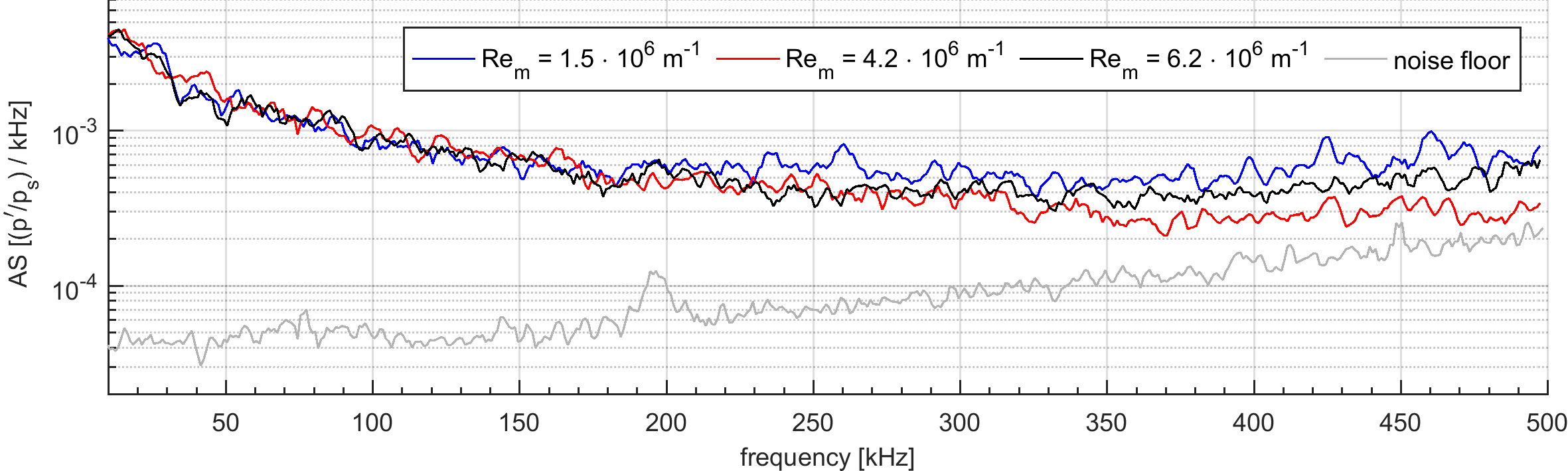}\label{fig:ASHEG}} \\
				\subfloat[HLB Mach 6]{\includegraphics[width=0.99\textwidth]{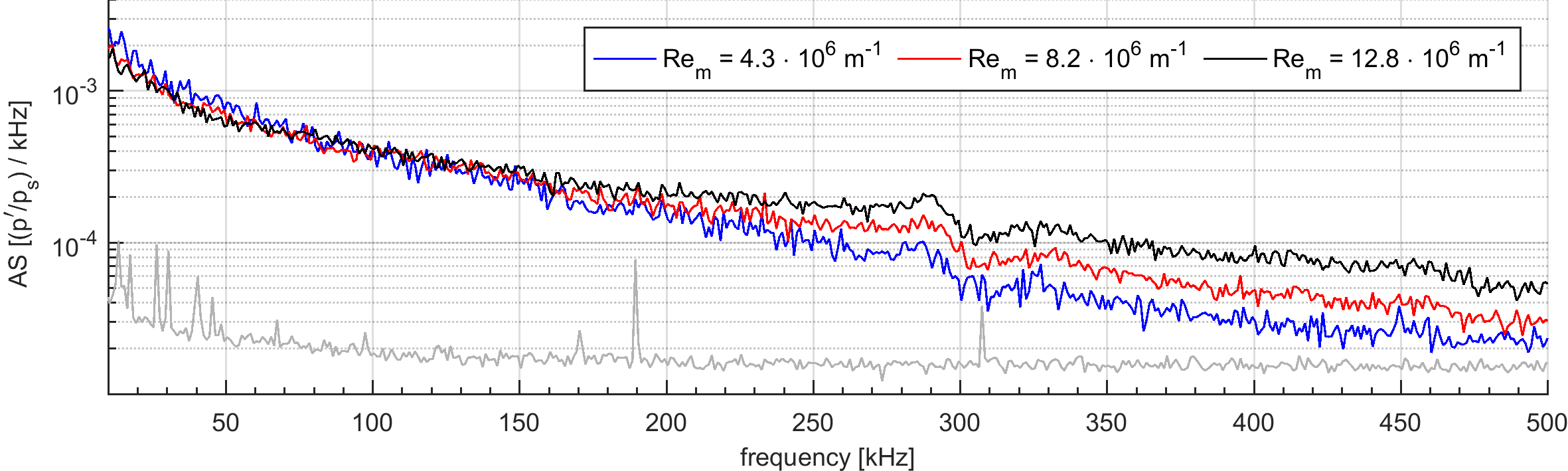}\label{fig:ASHLB}} \\
				\subfloat[RWG Mach 6]{\includegraphics[width=0.99\textwidth]{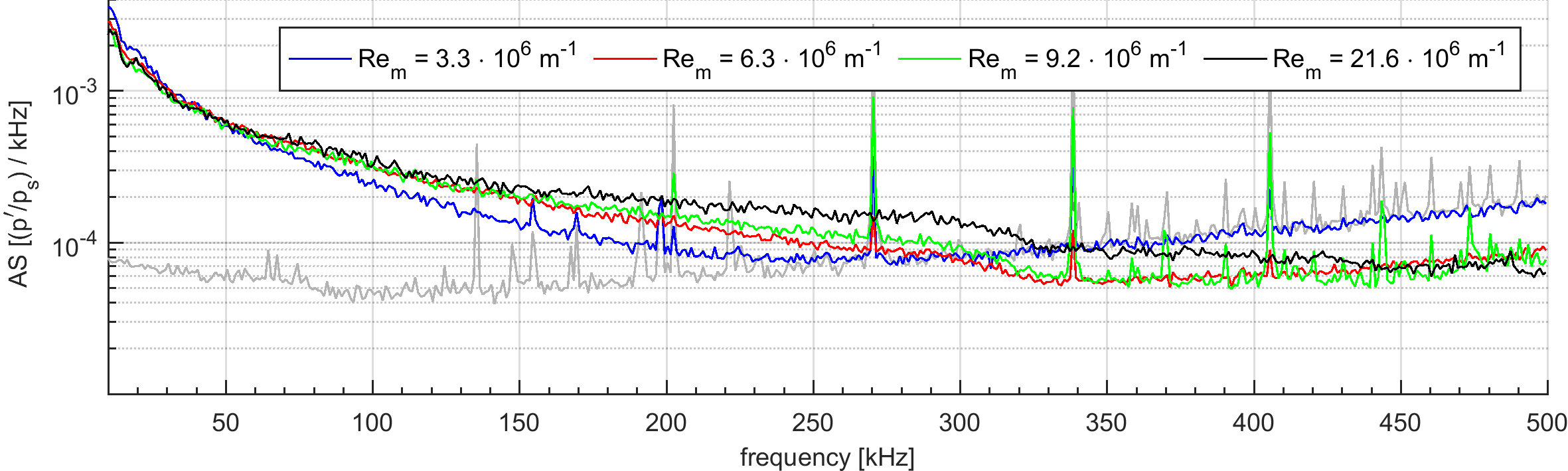}\label{fig:ASRWG}} \\
				\subfloat[RWG Mach 3]{\includegraphics[width=0.99\textwidth]{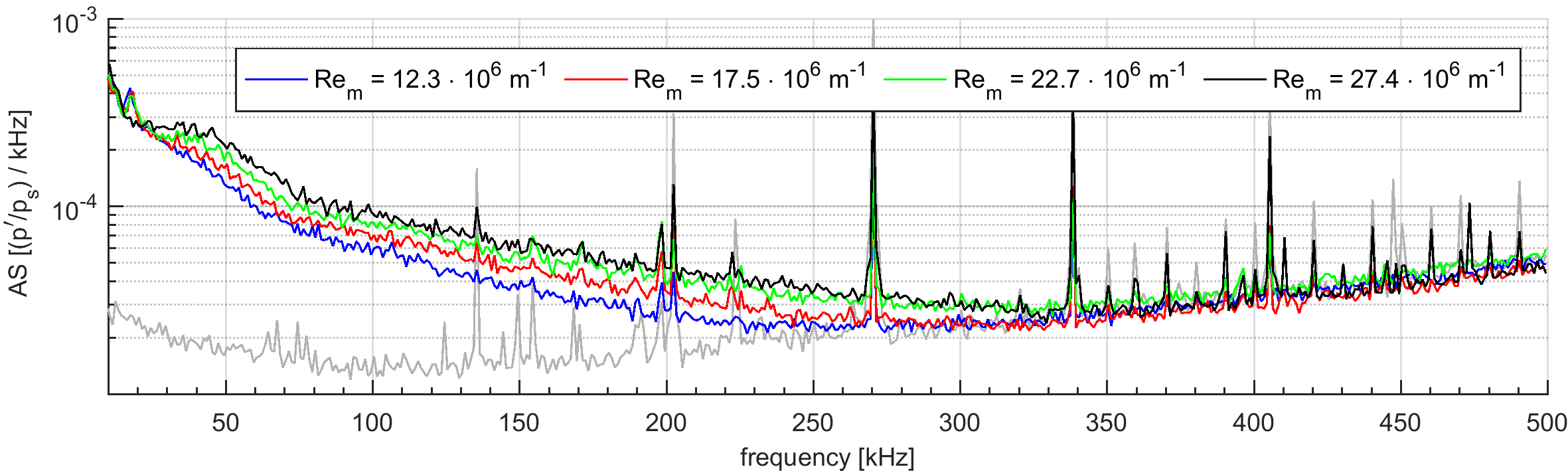}\label{fig:ASRWGM3}} \\
				\caption{Amplitude spectra (signal RMS in a \unit{1}{kHz} frequency window) of the piezoelectric transducer readings normalized to $\unit{1}{kHz}$ obtain in HEG, HLB and RWG using the wedge probe at Mach numbers 3, 6 and 7.4. The noise floor measured before each test is depicted in gray. The pressure readings were normalized using the measured probe surface pressure.}
				\label{fig:AS}
\end{figure}

Figures \ref{fig:ASHEG} to \ref{fig:ASRWGM3} depict the amplitude spectra (AS), computed as the RMS, of the normalized surface pressure based on a \unit{1}{kHz} interval according to equation \ref{equ:PStoRMS}. The signals were recorded in the three hypersonic facilities described in section \ref{sec:ExperimentalSetup} using flush mounted piezoelectric pressure transducers of type PCB132. To mechanically decouple the transducer from the probe, silicone sleeves were used around the transducers. For reasons of clarity only a subset of all flow conditions is plotted, representing the lowest, the highest and an intermediate unit Reynolds number in each facility. The frequency limit at which the signal reached the noise level varies between $\unit{\approx 300}{kHz}$ at Mach 3 in the RWG and $\unit{\approx 750}{kHz}$ at Mach 7.4 in HEG (not shown). The noise levels were found to be different in each facility. Apart from a small but repeatably visible bump in the HLB spectra at \unit{\approx 280}{kHz}, all spectra decay monotonically until the noise level is reached.  \\
Figures \ref{fig:ASHLB} and \ref{fig:ASRWG} allow a direct comparison between the HLB and RWG facilities, both of which are operated at similar test conditions with identical nozzle exit diameters. The obtained spectral distributions are found to be comparable, showing higher RMS values for lower unit Reynolds numbers below $\unit{\approx 50}{kHz}$ in RWG and below $\unit{\approx 100}{kHz}$ in HLB. At higher frequencies the signal amplitude is higher for larger unit Reynolds numbers, which indicates a shift of spectral energy towards higher frequencies. The tests at Mach 3 show a similar trend in a frequency range above $\unit{\approx 30}{kHz}$. The normalized pressure readings were found to be almost a magnitude lower compared to those obtained at higher Mach numbers. \\

Since the low frequency content of a signal strongly contributes to the RMS of a signal, an adequate low frequency limit needs to be found to provide a representative RMS for a short duration test facility such as HEG. In the present study a low frequency limit of \unit{1}{kHz} was chosen, corresponding to a disturbance time period of \unit{1}{ms}. We assume that disturbance frequencies below this limit are not of relevance with respect to the transition process driven by e.g. second mode instabilities at frequencies of the order of several \unit{100}{kHz}. Furthermore, it was found that a frequency limit of \unit{1}{kHz} is appropriate to obtain representative measures for test times in the range of a few milliseconds. 
\begin{figure}[htbp]
	\centering
    	\subfloat[\unit{\leq 50}{kHz}]{\includegraphics[width=0.99\textwidth]{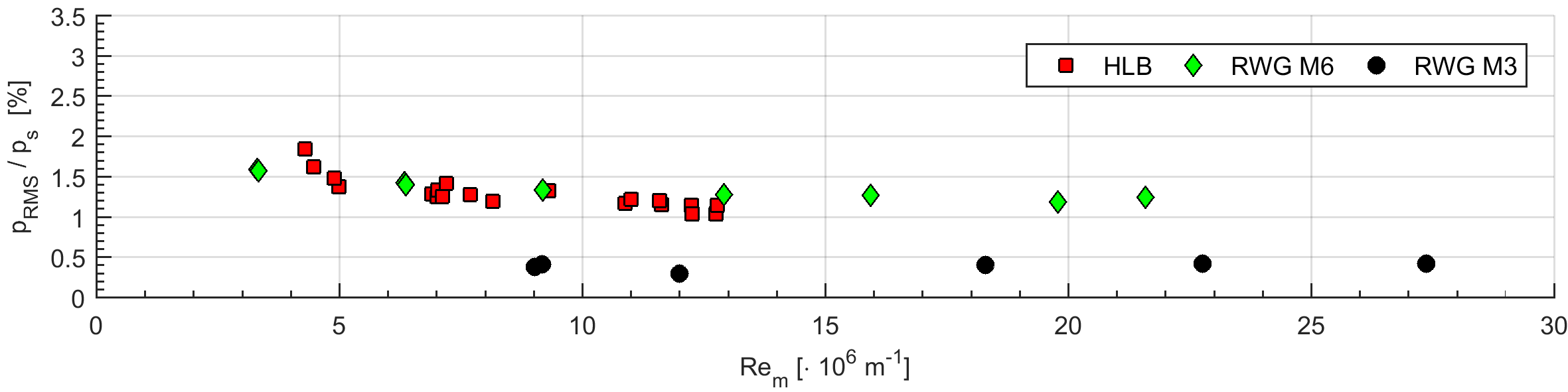}\label{fig:KPCBKU_intRMS0}} \\
		\subfloat[\unit{1 - 50}{kHz}]{\includegraphics[width=0.99\textwidth]{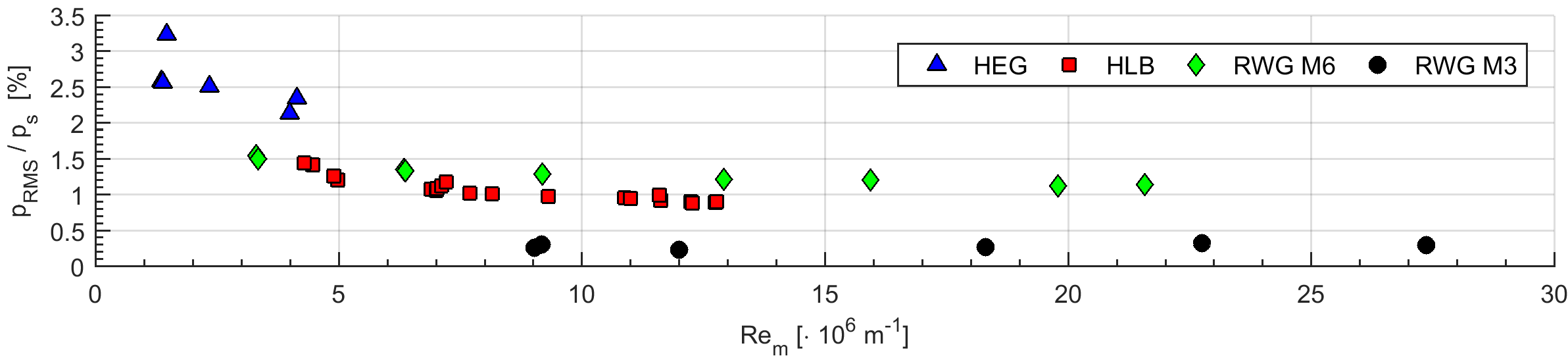}\label{fig:KPCBKU_intRMS1}} \\
		\subfloat[\unit{50 - 100}{kHz}]{\includegraphics[width=0.99\textwidth]{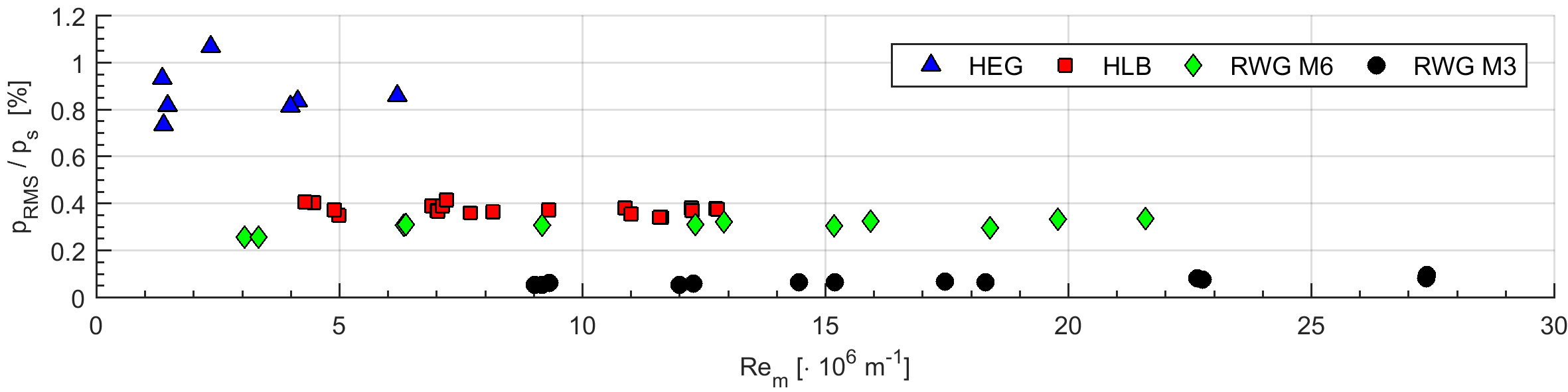}\label{fig:KPCBKU_intRMS2}} \\
		\subfloat[\unit{100 - 200}{kHz}]{\includegraphics[width=0.99\textwidth]{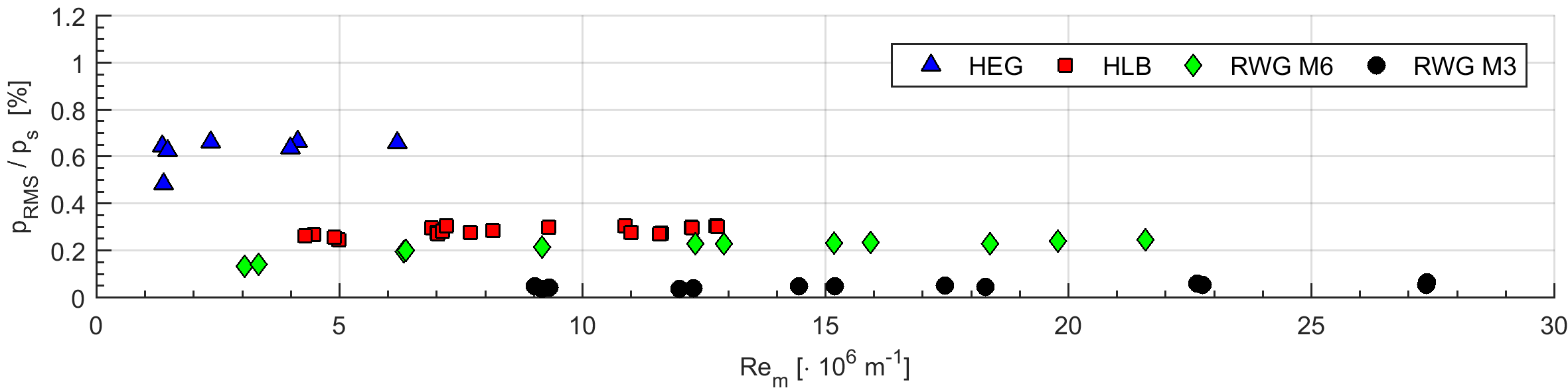}\label{fig:KPCBKU_intRMS3}} \\
		\subfloat[\unit{200 - 300}{kHz}]{\includegraphics[width=0.99\textwidth]{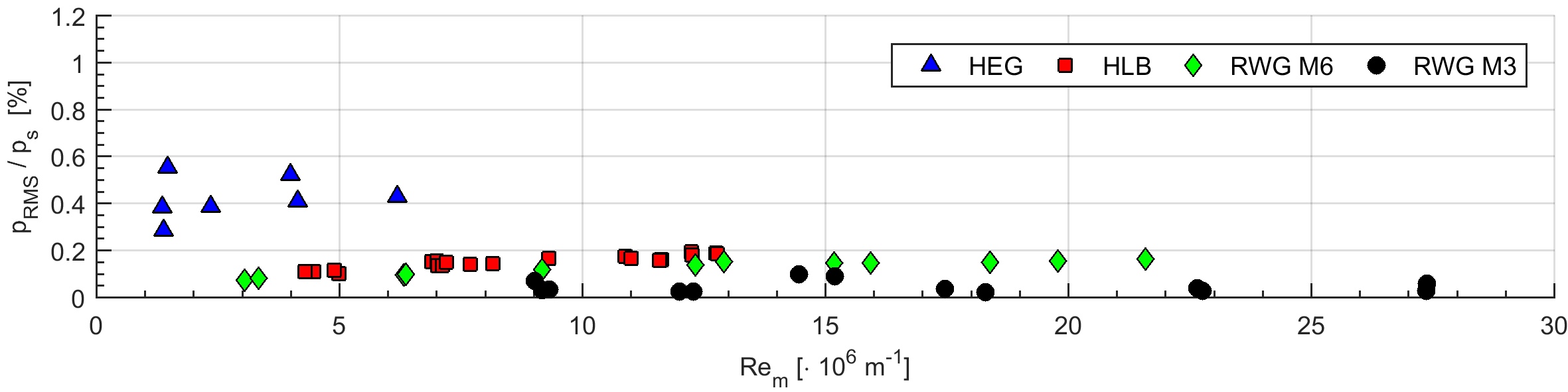}\label{fig:KPCBKU_intRMS4}} \\
	\caption{Surface pressure RMS normalized by mean surface pressure evaluated in five different frequency ranges.}
	\label{fig:KPCBKU_intRMS}
\end{figure}

In figure \ref{fig:KPCBKU_intRMS} the amplitude spectra measured over a wide unit Reynolds number range were evaluated in five different frequency ranges, providing surface pressure RMS estimations for each frequency range. The surface pressure measurements were normalized with the mean surface pressure on the probe. Figure \ref{fig:KPCBKU_intRMS0} provides the surface pressure RMS obtained in HLB and RWG up to a frequency of $\unit{50}{kHz}$. The lower frequency limit corresponds to the frequency resolution which is in the order of \unit{10}{Hz} depending on the available test time which depends on the facility and the test condition. The RMS values are mostly dominated by the low frequency disturbances and correspond to results often obtained by means of pitot probes. Figure \ref{fig:KPCBKU_intRMS1} provides the surface pressure RMS for all facilities including HEG in a frequency range of \unit{1}{kHz} to \unit{50}{kHz}. The increased low frequency limit leads to a decrease of the RMS levels and reduces the scatter of the data since low frequency, high amplitude events contribute less to the total RMS. The effect is observable in the RWG and the HLB data but is more pronounced in HLB and in RWG at Mach 3. The latter observation underlines the importance of the frequency limits to provide comparable RMS values.
In general, it can be seen that all RMS values in the frequency range decrease with increasing unit Reynolds number, except those obtained at Mach 3. The latter remain almost constant over the available unit Reynolds number range. At Mach 6 pressure RMS levels between \unit{1.0-1.8}{\%} and \unit{1.2-1.6}{\%} were measured in HLB and RWG. At Mach 3 pressure RMS levels in the range of \unit{0.3-0.4}{\%} were obtained in the latter facility. The highest levels were measured in HEG at Mach 7.4 with RMS levels between \unit{2.2-3.4}{\%}.
As expected, the subsequent figures, figure \ref{fig:KPCBKU_intRMS2} to \ref{fig:KPCBKU_intRMS4}, show distinctly lower RMS values compared to the low frequency range. Furthermore, the scatter of the data reduces since the high frequency disturbances are statistically better represented in the available test times. In contrast to figure \ref{fig:KPCBKU_intRMS1}, the trend of the RMS with increasing Reynolds number is reversed. It can be seen that, for instance, in figure \ref{fig:KPCBKU_intRMS4} the RMS increases with increasing Reynolds number. That also holds for the Mach 3 data and is observable in figure \ref{fig:KPCBKU_intRMS2} and \ref{fig:KPCBKU_intRMS3}. \\

The present approach provides access to the RMS levels in a specific frequency range of interest. In this way frequency-specific RMS values could, for instance, be used to support the study of the hypersonic boundary layer transition processes dominated by second mode instabilities. Previous transition studies in HEG revealed the second mode frequencies to be in a frequency range of \unit{250-500}{kHz} on a \unit{7}{\degree} half angle cone at low enthalpy conditions \cite{Wagner2014,Laurence2014}. According to the present study, the integrated surface pressure RMS level measured in a frequency range of \unit{250-350}{kHz} is \unit{0.4 \pm{0.1}}{\%}. Assuming a frequency range of \unit{200-300}{kHz} to be of interest for hypersonic transition studies in both Ludwieg tubes (Heitmann et al. \cite{Heitmann2011a}), the surface pressure RMS level in this frequency range is \unit{0.1 - 0.2}{\%} in HLB and \unit{0.07 - 0.16}{\%} in RWG over the full unit Reynolds number range of each facility as depicted in figure \ref{fig:KPCBKU_intRMS4}.\\

Considering the popularity of pitot probe measurements and the large amount of publicly-available data, the wedge probe results are compared to pitot probe measurements conducted simultaneously together with the wedge probe in HEG and RWG. In the latter the transducer was directly exposed to the flow while in HEG a half sphere pitot probe with a radius of \unit{7.5}{mm} was used in combination with a stretched cavity in front of the transducer to minimize the risk of particle impact or overheating. Due to the screen and the involved cavities, it is expected that the pitot probe setup causes damping of the high frequency content of the signal.

Stainback et al. \cite{Stainback1972} and Harvey et al. \cite{Harvey1969} reported a method to convert static pressure fluctuations, $p_{rms} / \bar{p}$, derived using a hot-wire anemometer, to total pressure fluctuations behind a normal shock, $p_{t,rms} / \bar{p}_t$. The method is based on the following relation between pitot pressure and dynamic pressure, valid above M=2.5 :\cite{Stainback1972}
\begin{equation}
p_{t,2} = G \rho u^2 \ ,
\end{equation}
where G is a constant depending only on $\gamma$. Assuming the equation to apply to instantaneous pitot pressures, and taking into account reflections of the compression waves at the pitot probe surface, the following relation between RMS pitot pressure fluctuations and the RMS of static pressure fluctuations was derived for plane moving sound waves:
	\begin{equation}
		\frac{p_{rms}}{\bar{p}} = \frac{\gamma}{2} \left( \frac{p_{t,rms}}{\bar{p}_t} \right) \left[ 1 - \frac{4 n_x}{M} + 4 \left( \frac{n_x}{M} \right)^2 \right]^{-1/2} \ ,
		\label{equ:PitotConversion}
	\end{equation}
with $n_x$ given  by
	\begin{equation}
		n_x = \left( \frac{u_s - u_{\infty}}{u_{\infty}}  \right)^{-1} M^{-1} \quad \textrm{and} \quad \frac{u_s}{u_{\infty}} = 0.6 \ , 
	\end{equation}
where $u_s$ is the sound source velocity.\cite{Stainback1972} 

\begin{figure}[htbp]
	\centering
	\includegraphics[width=0.60\textwidth]{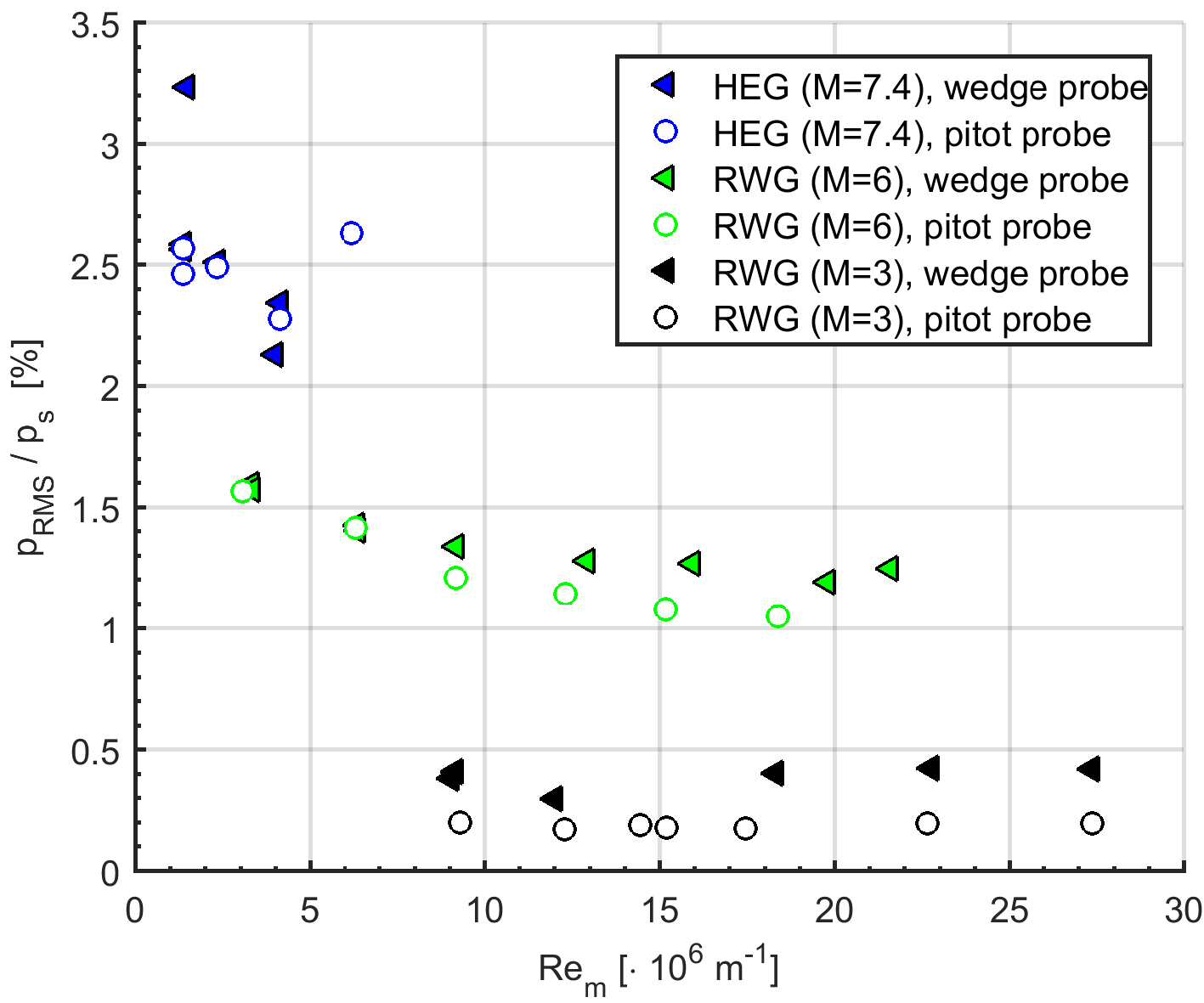}
	\caption{Comparison of pressure fluctuation measurements on the wedge probe with converted pitot probe pressure fluctuations using equation \ref{equ:PitotConversion}. The data was normalized with the mean surface pressure measured on the according probe surface.}
	\label{fig:PitotComparison}
\end{figure}

Figure \ref{fig:PitotComparison} provides the RMS of the pressure fluctuations measured on the wedge probe in HEG and RWG for frequencies up to $\unit{50}{kHz}$ and over a wide unit Reynolds number range. The RMS of the pitot pressure fluctuations obtained in both facilities were converted into according static pressure fluctuations using equation \ref{equ:PitotConversion} and depicted in the same figure, with $\gamma=1.4$ assumed for all cases. The pressure readings were normalized with the mean surface pressure measured on each probe.   

It can be seen that for Mach 7.4 and Mach 6 the converted RMS pitot fluctuations match the wedge probe results well. However, the trend of the pitot pressure fluctuations with unit Reynolds number seems to be slightly different for the Mach 6 case, i.e. the pitot measurements indicate a faster drop of RMS with increasing Reynolds number, probably caused by damping mechanisms due to the pitot setup. At Mach 3 the converted pitot data are found to be about a factor of 2 lower compared to the RMS obtained on the wedge probe. The reason for the latter observation is presently unknown. The good agreement of the data for Mach numbers above 6 justifies the use of the wedge probe in test environments not accessible with pitot probes. It can be concluded that the proposed probe can be used in a wide range of hypersonic facilities to assess the static pressure fluctuations or to derive a pitot probe RMS equivalent.  \\  

\begin{figure}[htbp]
	\centering
		\includegraphics[width=1.00\textwidth]{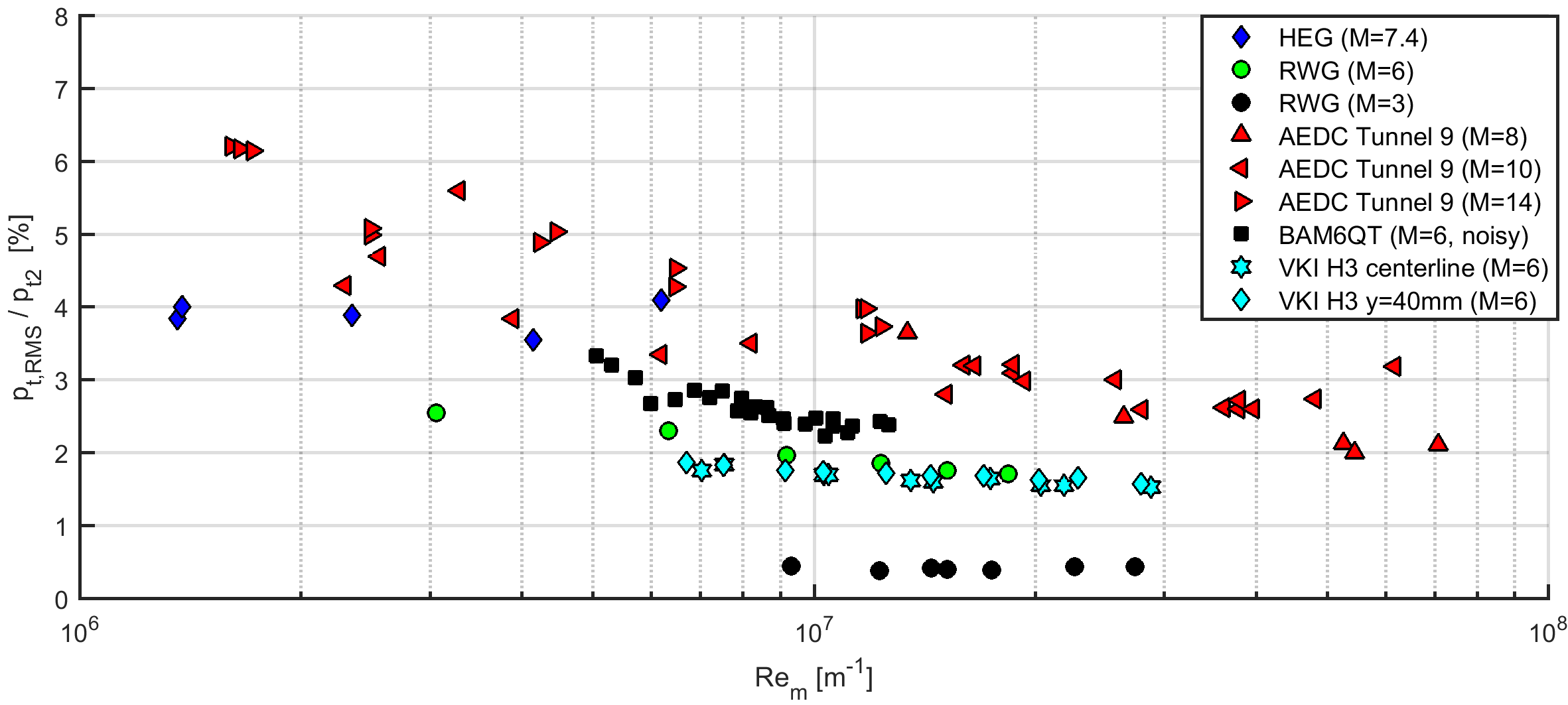}
		\caption{Normalized RMS pitot pressure fluctuations obtained in HEG and RWG combined with the corresponding data reconstructed from Masutti et al. \cite{Masutti2012}, Lafferty et al. \cite{Lafferty2007}, Steen \cite{Steen2010} and Rufer et al. \cite{Rufer2012}.}
	\label{fig:RMS_PITOT_ExtTunnels}
\end{figure}

Finally, figure \ref{fig:RMS_PITOT_ExtTunnels} relates the RMS pitot pressure fluctuations obtained in HEG and RWG to data obtained at comparable test conditions in the AEDC Tunnel 9 (Lafferty et al. \cite{Lafferty2007}), the VKI H3 (Masutti et al. \cite{Masutti2012}), the BAM6QT at noisy conditions (Steen \cite{Steen2010}) and the NASA Langley 20-Inch Mach 6 Wind Tunnel (Rufer et al. \cite{Rufer2012}). It can be seen that the RWG Mach 6 data, ranging from $\unit{1.7}{\%}$ to $\unit{2.6}{\%}$, show a good agreement with the RMS pitot fluctuations obtained in the VKI H3 facility at Mach 6. The RWG Mach 3 data range from $\unit{0.39}{\%}$ to $\unit{0.45}{\%}$ and thus remain well below all RMS levels obtained at higher Mach numbers. The observation is consistent with the results provided in figure \ref{fig:AS} and \ref{fig:KPCBKU_intRMS} as well as the general trend expected with decreasing free-stream Mach number. The RMS pitot pressure fluctuations measured in HEG scatter around $\unit{4}{\%}$ and do not show a clear trend with increasing Reynolds number.  

\section{Numerical Results}
\label{sec:numresults}

The aim of the present simulations is to provide transfer functions, namely the wall pressure fluctuation to free-stream pressure fluctuation ratio, along the wall and in particular at the pressure transducer locations used in the wind-tunnel experiments. 
\begin{figure}[htbp]
		\centering
		\subfloat[fast acoustic waves]{\includegraphics[width=0.49\columnwidth]{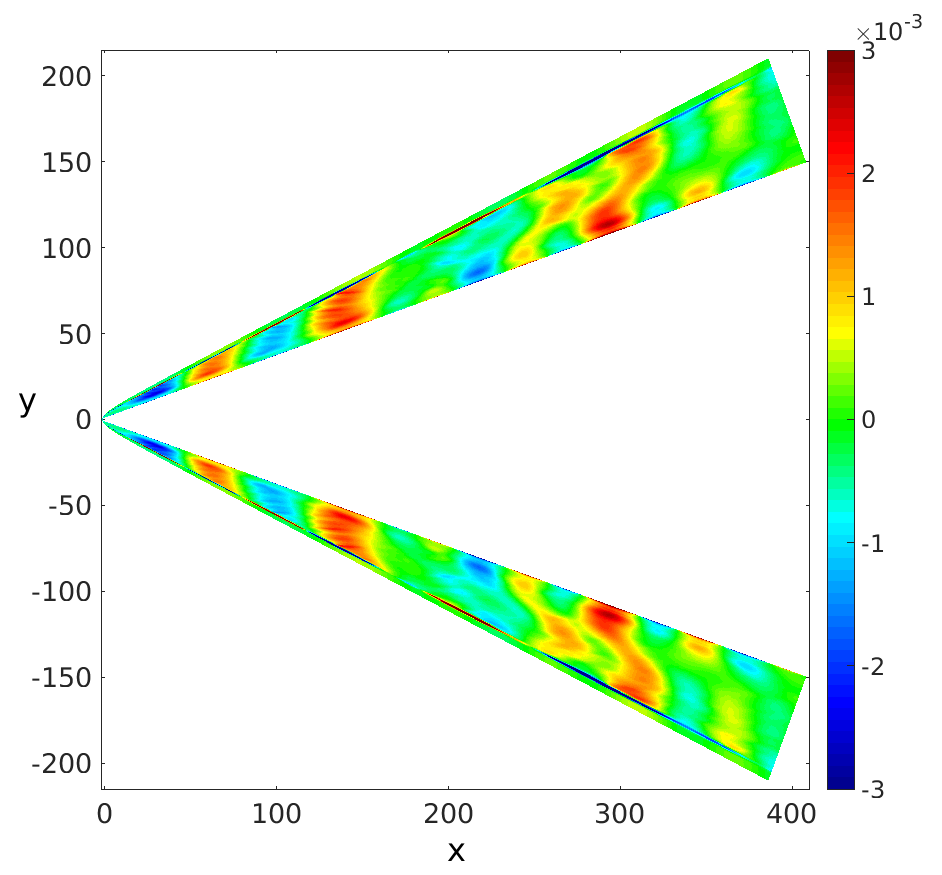} \label{fig:density_field}}
		\subfloat[slow acoustic waves]{\includegraphics[width=0.49\columnwidth]{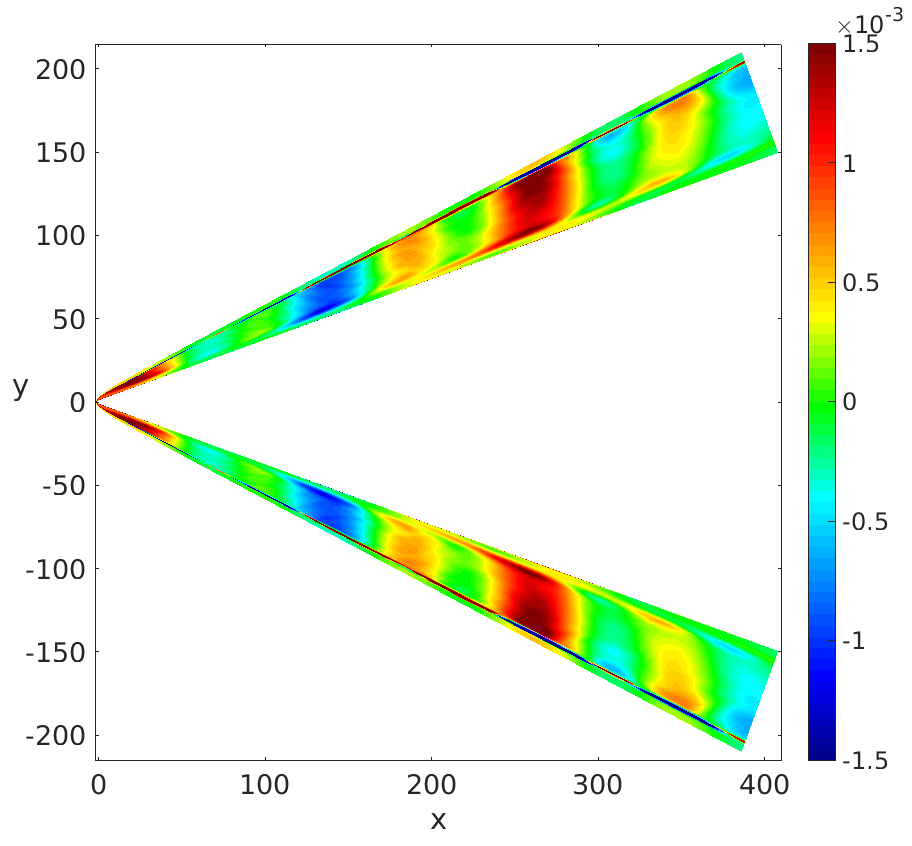} \label{fig:density_fieldslow}}
		\caption{Instantaneous density fluctuation field for Case 2, Mach=7.3, $Re_{m}$=\unit{1.4\times10^6}{m^{-1}}. All dimensions are normalized with a nose radius of \unit{0.1}{mm}.}
\end{figure}
Figures \ref{fig:density_field} and \ref{fig:density_fieldslow} illustrate the density disturbance field generated by a fast and slow acoustic wave transmitted in the post-shock region of test case 2 in table \ref{tab:flow_conditions}. The wave structure behind the shock is irregular due to the multiple frequencies that are inserted with random phases. In addition, the post-shock wave fronts are no longer vertical, since the wave fronts are affected by the change in convection speed across the shock.

For slow acoustic waves (figure \ref{fig:density_fieldslow}), the post-shock waves appear more compact, showing a higher wavelength and a lower deflection angle (i.e.~more aligned with the vertical axis), due respectively to the lower considered frequency range and the change in phase speed across the shock. Similar patterns have been observed for the other cases listed in table \ref{tab:flow_conditions}.

\begin{figure}[htbp]
		\centering
			\subfloat[fast acoustic waves]{\includegraphics[width=0.49\columnwidth]{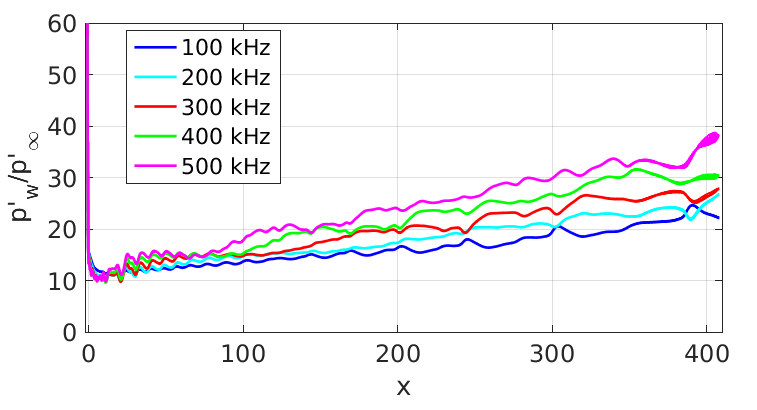} \label{fig:pw_distribution}} 
			\subfloat[slow acoustic waves]{\includegraphics[width=0.49\columnwidth]{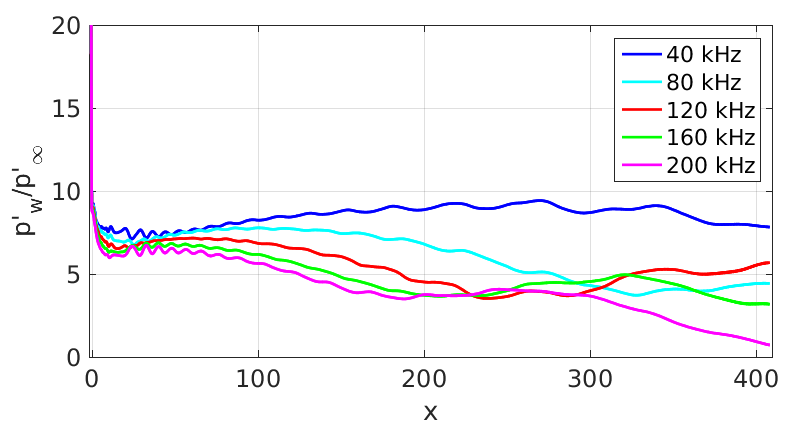} \label{fig:pw_distribution_slow}}
		\caption{Pressure fluctuation amplitude distribution along the wall at different frequencies for Case 1. Mach=7.3, $Re_{m}$=\unit{4.4\times10^{6}}{m^{-1}}. Note that the frequency ranges were chosen differently for the slow modes.}
\end{figure}

For Case 1, with fast and slow waves respectively, Figures \ref{fig:pw_distribution} and \ref{fig:pw_distribution_slow} show the distribution of the wall pressure fluctuation amplitudes (normalized with the pressure fluctuation amplitude in the free-stream) for five different frequencies, computed through a Fast Fourier Transform approach. 
It should be mentioned that the relationship between the Fourier transformed pressure fluctuation amplitude ($p'$), which is the value we refer to in the present section, and the corresponding RMS value ($p'_{RMS}$), at each single frequency, is $p'=\sqrt{2}\ p'_{RMS}$.  
For fast acoustic waves (figure \ref{fig:pw_distribution}) the fluctuation amplitude grows gradually with both distance from the leading edge and with frequency. This behavior is due to a frequency-dependent resonance (or synchronization) mechanism in the leading-edge region between the external fast acoustic waves and the induced boundary-layer fast mode (denoted as mode F by Fedorov \cite{Fedorov2010}), which has almost the same phase speed of the fast acoustic waves at the leading edge.

For slow acoustic waves (figure \ref{fig:pw_distribution_slow}), in contrast, the pressure fluctuation amplitude shows a frequency-dependent decreasing trend, with the exception of a weak early growth shown at 40 kHz, which precedes a decay further downstream. This behavior suggests a weaker resonance mechanism between the induced boundary-layer slow mode (mode S in Fedorov \cite{Fedorov2010}) and the forcing acoustic waves at the leading edge, compared to the case of fast acoustic waves. It should be noted that these results relate to the evolution of the wall response before any growth of the unstable second mode, which is not captured in our simulations as it would occur downstream of the computational domain. Hence, the numerical results show that the leading-edge region is characterized by an early amplification of the fast mode, due to a strong resonant interaction with the forcing fast acoustic waves, and an initial decay of the slow mode.

The frequency range for slow acoustic waves is lower compared to fast acoustic waves. These different frequency ranges were chosen in order to provide high frequency resolution (with a maximum of 10 frequencies) around the most interesting frequencies in terms of growth behavior of the induced boundary-layer modes. An important difference is observed between fast and slow acoustic waves relative to the trend of the wall response with the frequency, namely the higher the frequency the higher is the growth rate of the fast-mode amplitude, but the higher the decay rate of the slow-mode amplitude. Hence, while in the case of fast acoustic waves the highest amplitudes of the internal modes induced in the leading edge region are reached at higher frequencies, the highest amplitudes for slow waves are reached at the lower frequencies. The pronounced early decay of the slow-mode response in the leading-edge region at higher frequencies is a feature of cold wall cases (to which our considered case pertains), as shown by Kara et al. \cite{Kara2007a}.   

\begin{figure}[h]
		\centering
			\subfloat[Mach 6, Case 4]{\includegraphics[width=0.49\columnwidth]{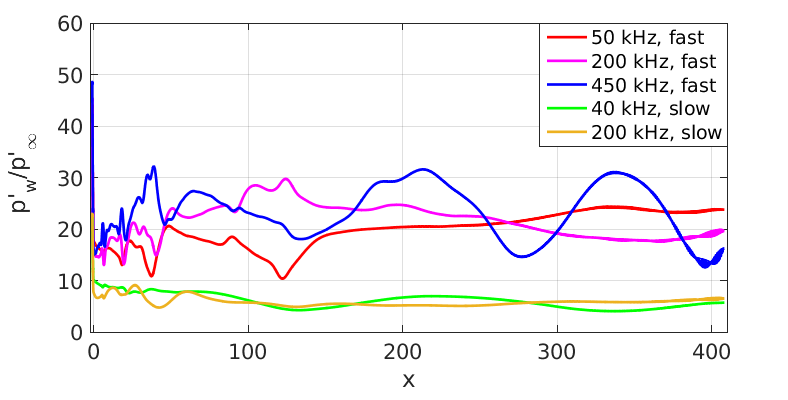} \label{fig:pw_distributionM6}} 
			\subfloat[Mach 3, Case 5]{\includegraphics[width=0.49\columnwidth]{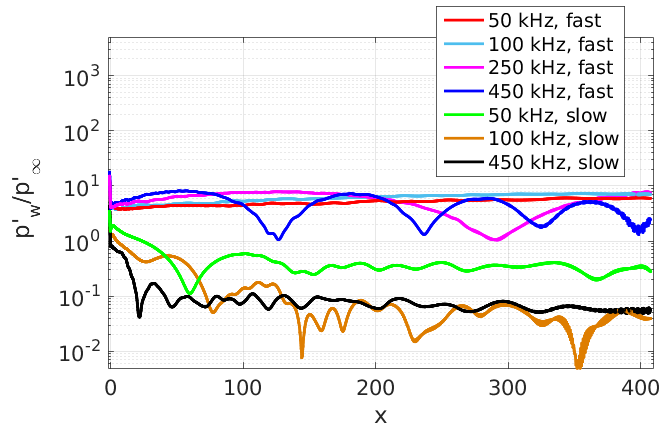} \label{fig:pw_distributionM3}}
		\caption{Pressure fluctuation amplitude distribution along the wall at the different frequencies. M=6: $Re_{m}$=\unit{6.3\times10^{6}}{m^{-1}}; M=3: $Re_{m}$=\unit{12.0\times10^{6}}{m^{-1}}}		
\end{figure}

Figure \ref{fig:pw_distributionM6} shows a comparison of the wall pressure fluctuations for Case 4 (Mach 6). In this case, the fast mode shows an oscillatory behavior at higher frequencies, representing another important feature of the receptivity to fast acoustic waves, namely a spatial modulation. This occurs downstream of the first peak (reached at $x \approx 40$ at 450 kHz for fast waves), and is caused by the fast mode being no longer synchronized with the forcing fast acoustic waves, as also seen in the work of Zhong and Ma \cite{zhong2006}.

In contrast, the response to slow acoustic waves in figure \ref{fig:pw_distributionM6} shows a lower-amplitude flatter trend at both the low (40 kHz) and the high (200 kHz) frequencies, which suggests that the slow mode, after an initial decay in the early nose region, remains at a substantially constant amplitude downstream. As with the fast mode, this region of constant amplitude for the slow mode precedes any unstable region that would occur downstream of the computational domain.

Figure \ref{fig:pw_distributionM3} shows the wall response for Case 5 (Mach 3). Here, the fast mode shows a pronounced oscillatory behavior at the higher frequencies, due to the modulation mechanism discussed above.

For slow acoustic waves, shown in the same figure, the wall response has a lower amplitude and a flat trend, with the exception of small-wavelength oscillations seen at 100 kHz. However, for this case the slow mode shows significantly higher amplitudes at the lowest frequency (50 kHz), compared to the response at the higher frequencies (100 kHz and 450 kHz). Note that the $y$-axis is presented in a logarithmic scale here, due to a high difference in the amplitude levels between the fast and the slow mode response.    

\begin{figure}[htpb]
		\centering
			\includegraphics[width=0.70\columnwidth]{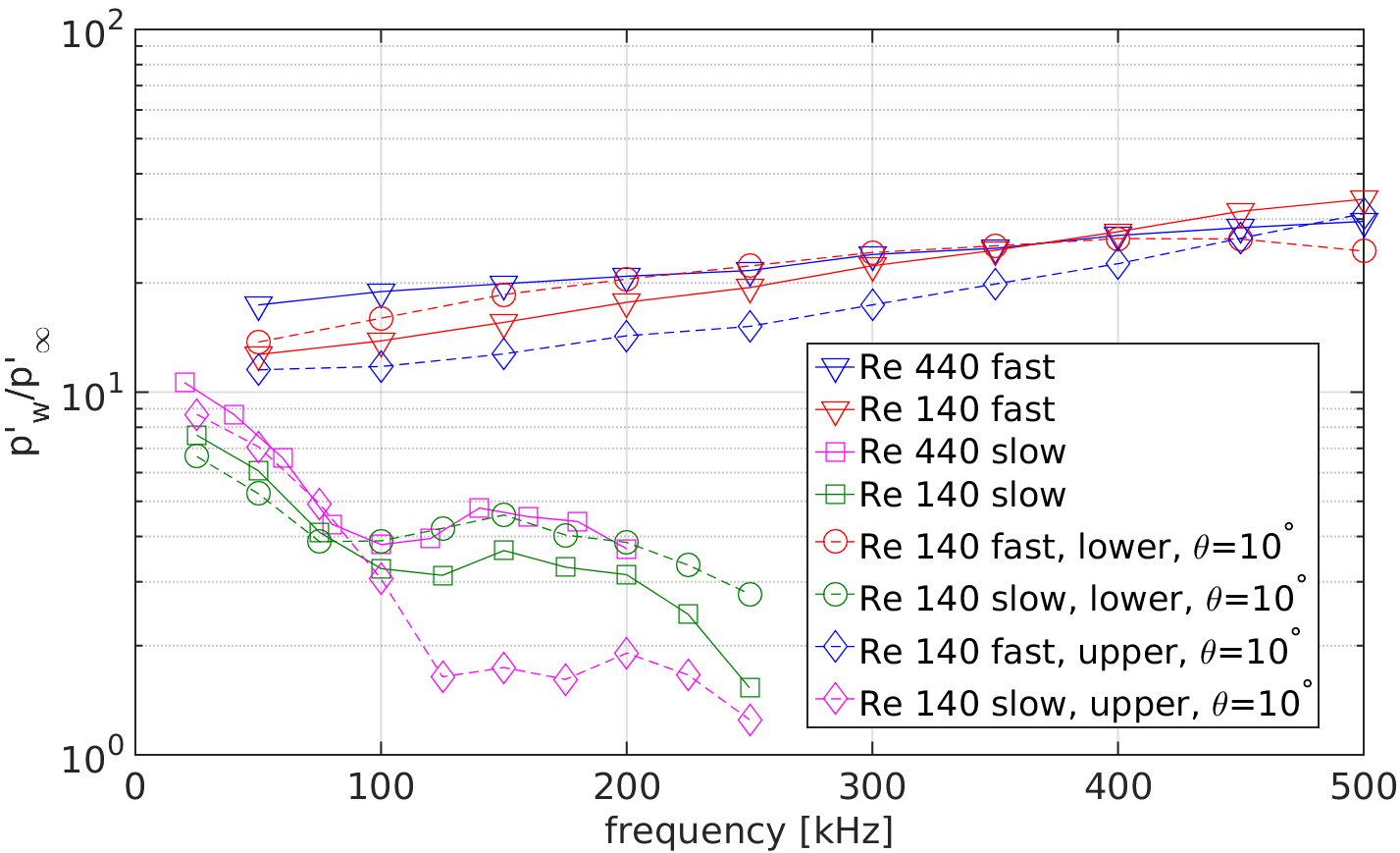}
		\caption{Frequency spectrum of the pressure fluctuation amplitudes at the transducer position ($x=297.3$) for Cases 1 to 3 (HEG) with both fast and slow acoustic waves. Mach=7.3, $Re_{m}$=\unit{4.4\times10^{6}}{m^{-1}} (Case 1), Mach=7.3, $Re_{m}$=\unit{1.4\times10^{6}}{m^{-1}} (Case 2 and 3). For Case 3, an inclination angle $\theta=10^{\circ}$ of the incident waves is considered.} 
%the response is presented on both the windward (bottom) side and the lee (upper) side of the wedge, due to the inclination angle $\theta=10^{\circ}$ of the acoustic waves with respect to the mean flow direction.}
		\label{fig:HEG_spectrum}
\end{figure}

To help interpret the experimental results, we now focus on the behavior at a transducer location ($x=297.3 $). Figure \ref{fig:HEG_spectrum} shows frequency spectra for Cases 1 to 3 (HEG), with fast and slow acoustic waves, in the frequency ranges described previously. It shows the effects of Reynolds number ($Re_{m}$=\unit{4.4\times10^{6}}{m^{-1}} for Case 1, $Re_{m}$=\unit{1.4\times10^{6}}{m^{-1}} for Cases 2 and 3) and angle of incidence of the acoustic waves ($\theta=10^{\circ}$ for Case 3). 
For Case 3 (Mach=7.3, $Re$=140, $\theta$=\unit{10}{\degree}) the results are shown on both the lower (wave-facing) and upper surfaces. 
As can be seen, for all the cases considered in figure \ref{fig:HEG_spectrum}, the results show a significantly higher response for fast waves than for slow waves at all frequencies. This is due to the stronger resonance mechanism at the leading edge, characterizing the receptivity to fast acoustic waves. The response to fast acoustic waves is seen to increase gradually with frequency, while the slow waves decrease in amplitude, with the minimum value being reached at the highest frequency. For fast acoustic waves, an angle of incidence of $10^{\circ}$ (the dashed curves in figure \ref{fig:HEG_spectrum}) is seen to produce a slightly higher response on the lower (wave-facing) side, compared to the case with zero incidence angle, and a slightly lower response on the upper side, except for frequencies higher than 450 kHz. For slow acoustic waves, the response on the upper side is higher and the response on the opposite side lower in the lower frequency range. However, for slow waves, at higher frequencies (higher than 100 kHz), a reversal is shown, namely the response on the lower side is significantly higher than on the upper side, due to a much more rapid decay of the amplitude at increasing frequencies observed on the upper side. The Reynolds number is seen to have little effect only on the amplitude of the wall response between cases 1 and 2 at all frequencies.

\begin{figure}[htpb]
\centering
\subfloat[HEG, Mach 7.3]{\includegraphics[width=0.60\columnwidth]{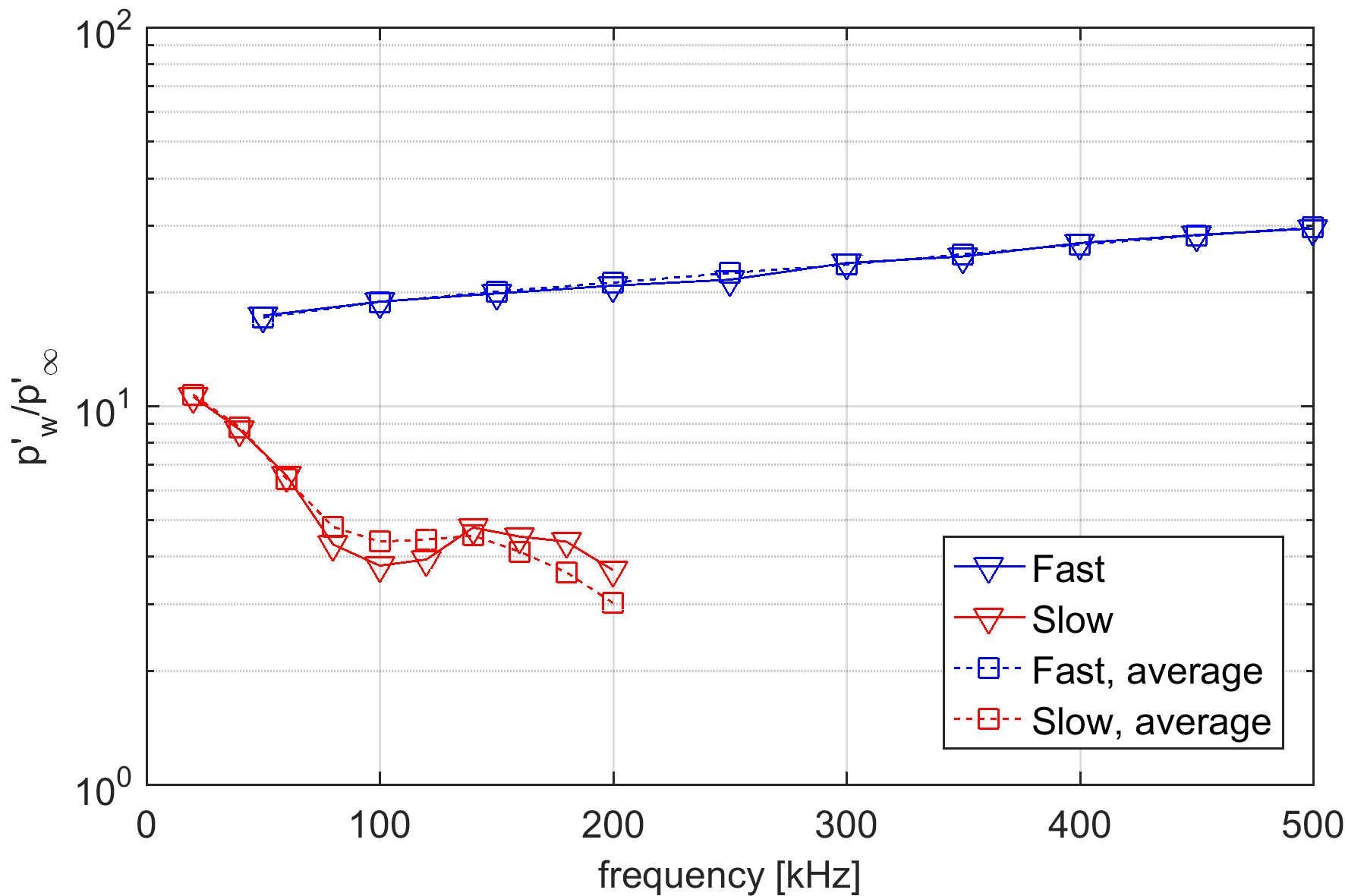}\label{fig:HEG_spectrum_case1}} \\
\subfloat[RWG, Mach 6]{\includegraphics[width=0.60\columnwidth]{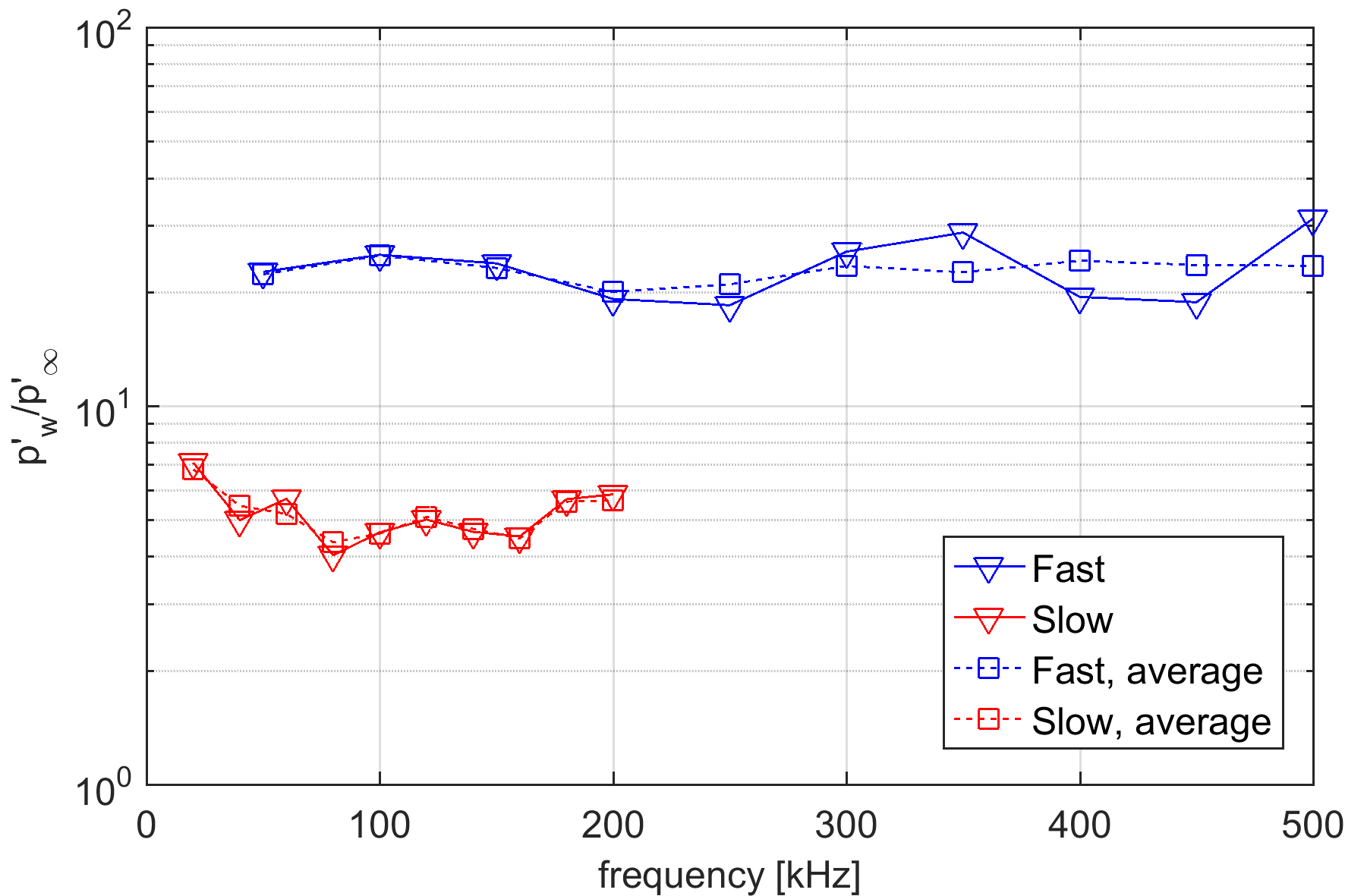}\label{fig:RWG_M6_spectrum}} \\
\subfloat[RWG, Mach 3]{\includegraphics[width=0.60\columnwidth]{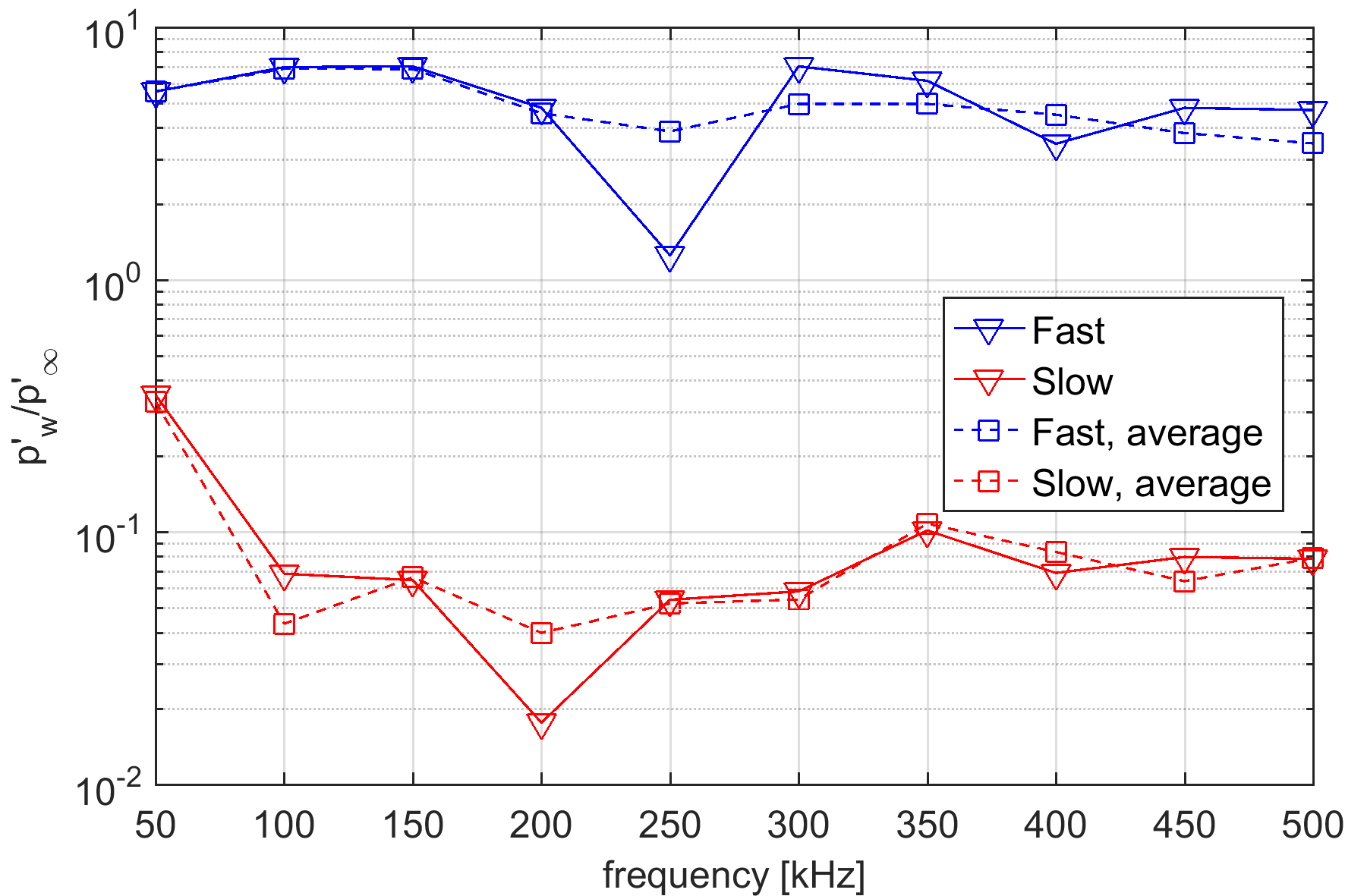}\label{fig:RWG_M3_spectrum}} \\
\caption{Frequency spectra of the pressure fluctuation amplitudes at the transducer position \unit{x=29.83}{mm} with both fast and slow acoustic waves, with and without spatial averaging.}
\label{fig:average_spectra}
\end{figure}

Figures \ref{fig:HEG_spectrum_case1} (Case 1), \ref{fig:RWG_M6_spectrum} (Case 4) and \ref{fig:RWG_M3_spectrum} (Case 5) show a comparison between the frequency spectra of the wall pressure fluctuation amplitude at the position $x=297.3$. Since the modulation behavior described earlier may lead to locally very low amplitudes at certain points on the surface, and these points would be expected to move around in experiments due to small variations in free stream conditions, the figures also include a spatial average over the region $x=200-400$. All the cases depicted in figure \ref{fig:average_spectra} show a significantly higher-amplitude response for fast acoustic waves, compared to slow waves.  Case 1, in figure \ref{fig:HEG_spectrum_case1}, shows the features already discussed in connection with figure \ref{fig:HEG_spectrum} and a very small effect of the space average on the frequency spectrum of the wall response.
For the RWG case at Mach 6 (Case 4, in figure \ref{fig:RWG_M6_spectrum}), the wall response shows a substantially flat trend with frequency for both fast and slow modes, as can be observed for the curves relative to the averaged spectra. In this case, the spatial averaging produces a slight change in the shape of the response for fast waves, and almost the same profile of the not averaged spectrum for slow waves.

The response for Case 5 (RWG, Mach 3, figure \ref{fig:RWG_M3_spectrum}), for both fast and slow acoustic waves, is significantly lower than for the higher Mach number cases. Additionally, the discrepancy in the amplitude between the fast and the slow mode response is much higher compared to the other cases. This clearly indicates an important Mach number effect on the receptivity to acoustic waves in supersonic flows. Moreover, in contrast to the other cases, the not averaged frequency spectrum for Mach 3 (Case 5) shows a local minimum, at 250 kHz for fast waves, and at 200 kHz for slow waves, compared to the average value of the amplitude level at the other frequencies. This is a consequence of the modulation of the pressure fluctuation response along the wall, as was shown in figure \ref{fig:pw_distributionM3} and described above.

\begin{figure}[htpb]
\centering        
\subfloat[HEG, Mach 7.3, $a_{F}=21.26$, $a_{S}=3.03$]{\includegraphics[width=0.60\columnwidth]{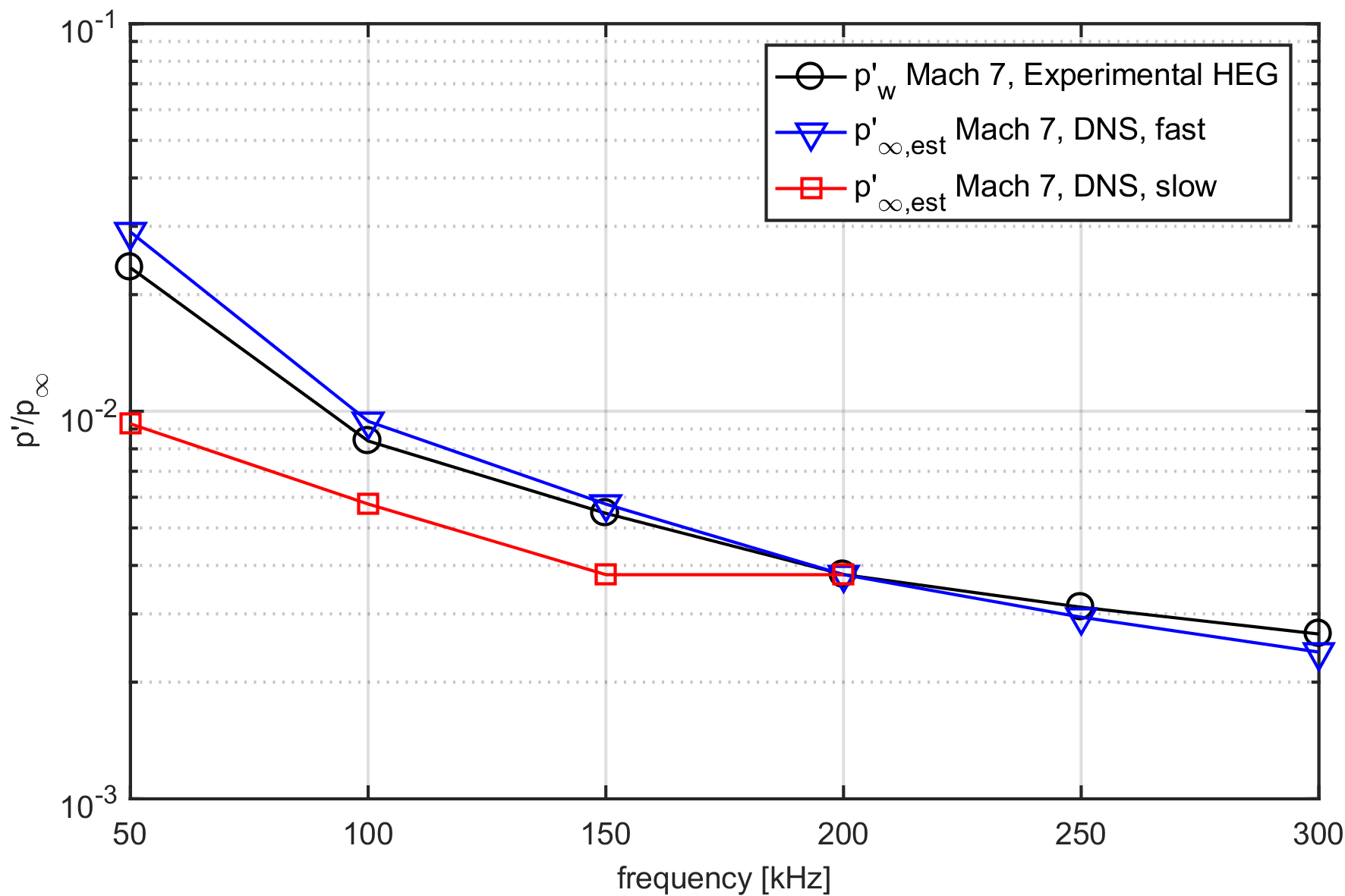}\label{fig:Estimated_levels_HEG_M7}} \\
%\caption{Integrated experimental wall pressure fluctuations over a window of 50 kHz (black solid line) for the HEG Mach 7 case, and estimated free-stream fluctuation amplitude levels for different disturbance combinations: 100\% fast, 100\% slow, 90\% fast-10\% slow, 10\% fast-90\% slow (dashed blue and red, and dotted blue and red curves respectively)}
\subfloat[RWG, Mach 6, $a_{F}=20.09$, $a_{S}=5.63$]{\includegraphics[width=0.60\columnwidth]{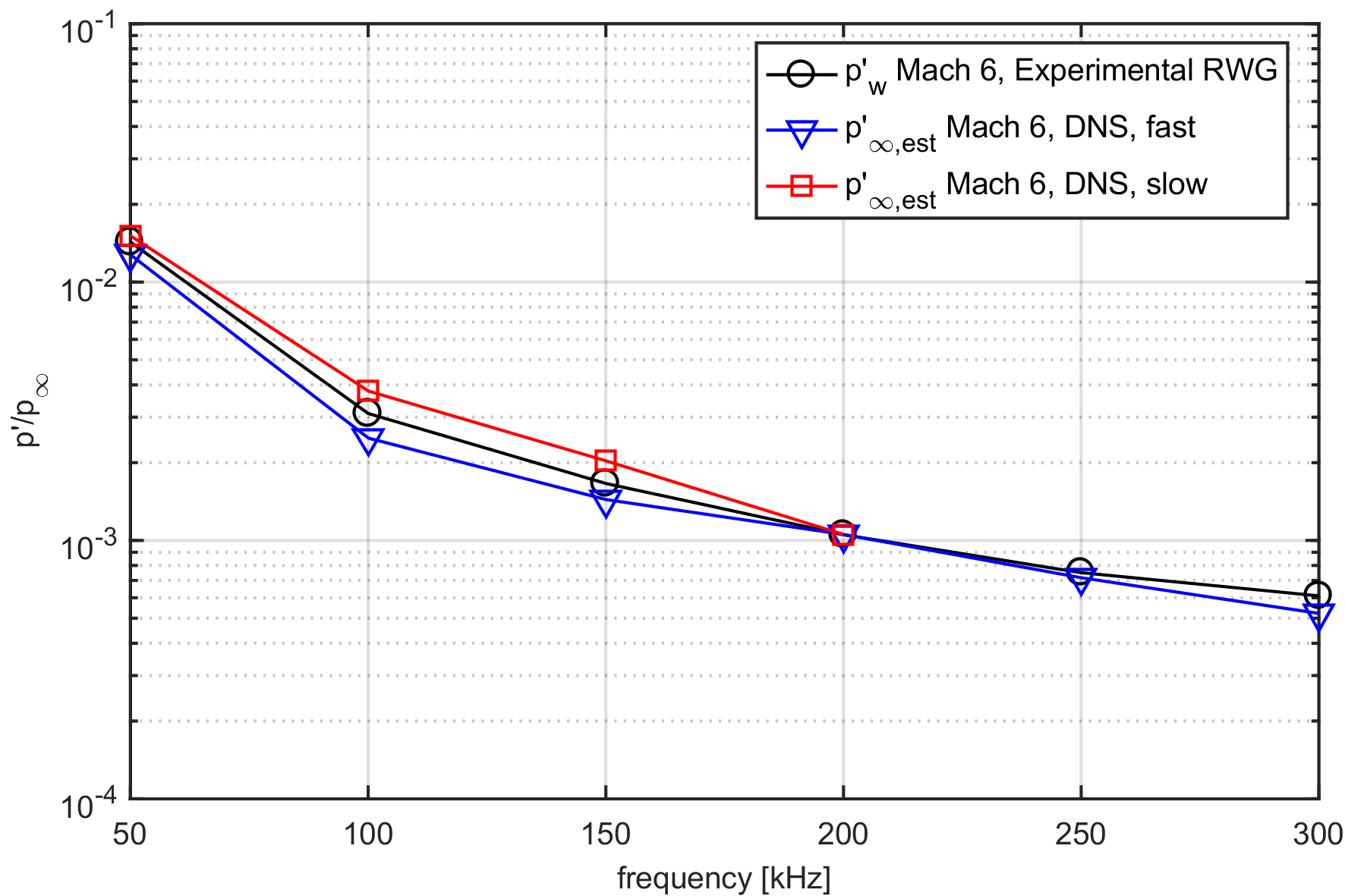}\label{fig:Estimated_levels_RWG_M6}} \\
%\caption{Integrated experimental wall pressure fluctuations over a window of 50 kHz (black solid line) for the RWG Mach 6 case, and estimated free-stream fluctuation amplitude levels for different disturbance combinations: 100\% fast, 100\% slow, 90\% fast-10\% slow, 10\% fast-90\% slow (dashed blue and red, and dotted blue and red curves respectively)}

\subfloat[RWG, Mach 3, $a_{F}=4.60$, $a_{S}=0.04$]{\includegraphics[width=0.60\columnwidth]{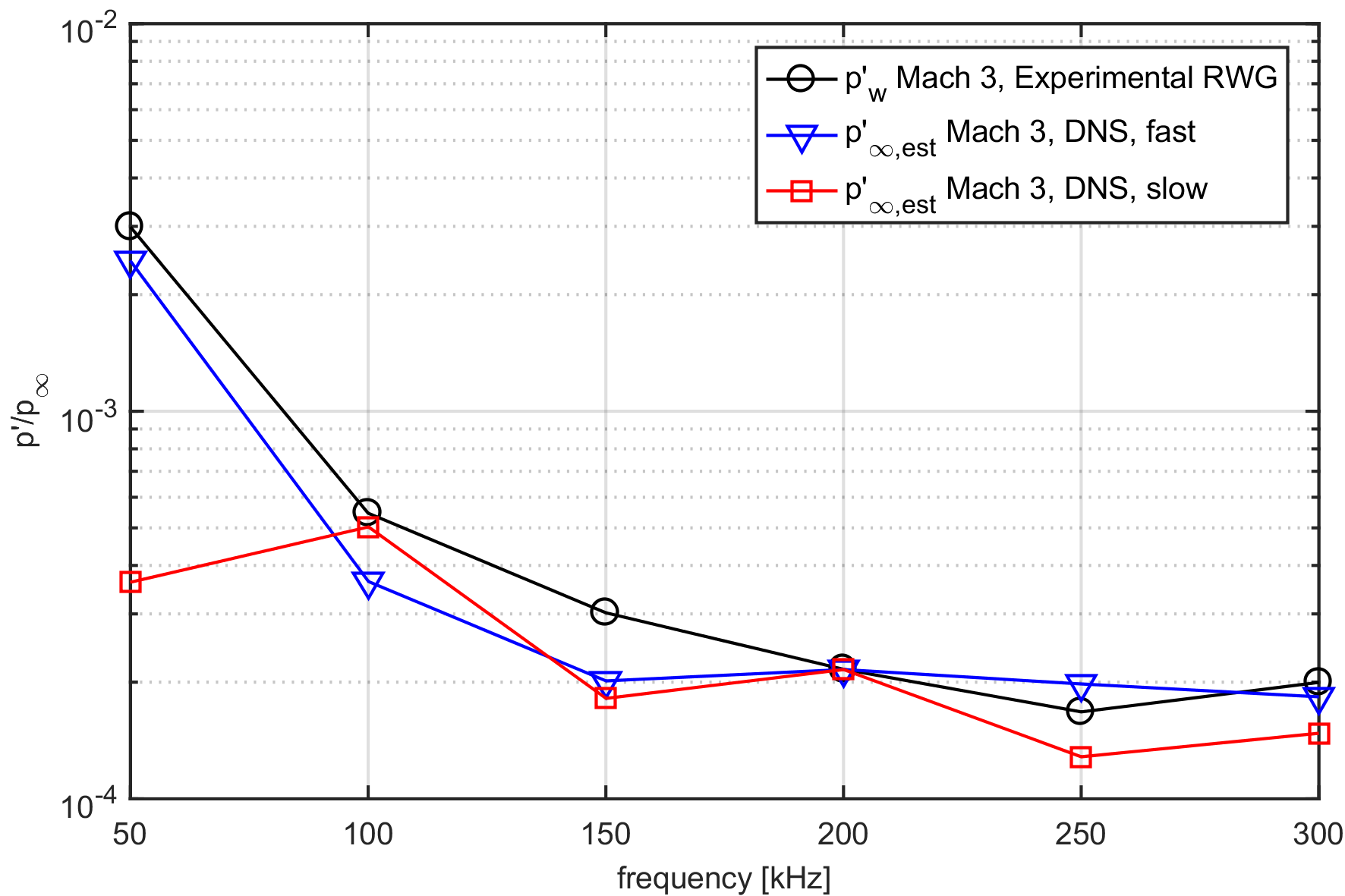}\label{fig:Estimated_levels_RWG_M3}} \\
\caption{Experimental data (integrated over a window of 50 kHz) and estimated free-stream noise levels, which have been anchored here to the experimental profiles at the reference frequency of 200 kHz through multiplication by the relative scaling factors $a_{F}$ (for fast waves) and $a_{S}$ (for slow waves).}
\label{fig:Wall_pressure_levels}
\end{figure}

The effect of receptivity can now be assessed by combining the numerical transfer functions with the wall pressure fluctuations measured in the experiments. This is done in two stages, first by comparing the effect on the spectral shape, which turns out to be small, and then considering the presence of either fast or slow modes in the free-stream. Figure \ref{fig:Wall_pressure_levels} shows the experimental frequency spectra of wall pressure fluctuation levels ($p'_{w}$), integrated over a window of \unit{50}{kHz}. The same frequency spectra were used to estimate the free-stream disturbance spectra using the numerical transfer functions ($p'_{\infty}/p'_{w}$), through the relation $p'_{\infty,est}=p'_{w,exp}(p'_{\infty}/p'_{w})$, for Case 1 (HEG Mach 7.3, $Re_{m}$=\unit{4.4\times10^{6}}{m^{-1}}), Case 4 (RWG Mach 6) and Case 5 (RWG Mach 3) respectively. The results presented in figure \ref{fig:Wall_pressure_levels} are normalized with the values of the free-stream mean pressure ($p_{\infty}^{*}$) related to each case, shown in Table \ref{tab:flow_conditions}. For each case, the transfer functions used to obtain the estimated noise levels in figure \ref{fig:Wall_pressure_levels} are the inverse functions of the corresponding wall-to-free-stream pressure fluctuation frequency spectra relative to the averaged response, shown in figure \ref{fig:average_spectra}. To better compare the shapes of the spectra, each curve of the estimated noise levels is anchored to the curve of the wall values through multiplication by a scaling factor computed at \unit{200}{kHz}. The corresponding scaling factors for fast and slow acoustic waves are indicated with the terms $a_F$ and $a_S$ respectively on the figures. For each case, the shape of the estimated noise level curves is very similar to the shape of the experimental curve, for both fast and slow acoustic waves, due to the flatness of the frequency responses shown in figure \ref{fig:average_spectra}. Exceptions need to be made for Case 1 with slow acoustic waves (figure \ref{fig:Estimated_levels_HEG_M7}) and Case 5 (Mach 3) for slow waves at the lowest frequency (50 kHz) in figure \ref{fig:Estimated_levels_RWG_M3}.

For all the cases in table \ref{tab:flow_conditions}, the scaling factors relative to the estimated noise levels for fast waves ($a_{F}=21.26, 20.09, 4.60$ for the Mach 7.3, Mach 6 and Mach 3 cases respectively) are significantly higher than the corresponding scaling factors for slow acoustic waves ($a_{S}=3.03, 5.63, 0.04$ for the Mach 7.3, Mach 6 and Mach 3 cases respectively). This means that a much lower noise level composed of only fast acoustic disturbances would be needed to provide the same experimental wall response, compared to the case of a free-stream composed by only slow acoustic waves. Moreover, the difference in the scaling factors between fast and slow acoustic waves is significantly higher for the Mach 3 case (Case 5), with $a_{F}$ being higher than $a_{s}$ by two orders of magnitude, which shows that there are strong Mach number effects.

%To investigate these effects, two arbitrary free-stream wave combinations, namely $90\%$ fast - $10\%$ slow and $10\%$ fast - $90\%$ slow, are considered. 
In order to provide a more reliable measure of the estimated noise levels for a particular frequency, coming from the combined experimental wall pressure values and the numerical transfer functions, the experimental RMS values were recomputed over a frequency window of 1 kHz between the frequencies 100 kHz and 101 kHz, with the numerical transfer functions at the frequency 100 kHz applied to obtain an estimation of the free-stream noise levels. The result is reported in table \ref{tab:noise_levels}, which shows, for Cases 1, 4 and 5, the experimental RMS values in the window 100-101 kHz ($(p'_{w}/p_{\infty})_{Exp}$), the numerical transfer functions for both fast and slow acoustic waves (denoted by $TF_{F}$ and $TF_{S}$ respectively) at the frequency 100 kHz, and the corresponding estimated noise levels for fast and slow waves ($(p'_{\infty}/p_{\infty})_{F}$ and $(p'_{\infty}/p_{\infty})_{S}$ respectively). Note that the numerical transfer functions correspond to the inverse functions of the values in the frequency spectra in figure \ref{fig:average_spectra}, namely $p'_{\infty}/p'_{w}$. 

\begin{table*}[htbp]
 \begin{center}
  \begin{tabular}{l|c|c|c|c|c}
       Facility & $(p'_{w}/p_{\infty})_{Exp}$ & $TF_{F}$ & $TF_{S}$ &  $(p'_{\infty}/p_{\infty})_{F}$ & $(p'_{\infty}/p_{\infty})_{S}$ \\\hline
      HEG Mach 7.3 (Case 1) & $8.9 \times 10^{-4}$ & $5.3 \times 10^{-2}$ & $2.27\times 10^{-1}$ & $4.7\times 10^{-5}$ & $2.0\times 10^{-4}$ \\
      RWG Mach 6 (Case 4) & $3.0\times 10^{-4}$ & $4.0\times 10^{-2}$ & $2.17\times 10^{-1}$ & $1.2\times 10^{-5}$ & $6.5\times 10^{-5}$ \\
      RWG Mach 3 (Case 5) & $5.44\times 10^{-5}$ & $1.4\times 10^{-1}$ & $23.25\times 10^{0}$ & $7.6\times 10^{-6}$ & $1.26\times 10^{-3}$ \\
  \end{tabular}
   \caption{Estimated free-stream noise levels for fast (F) and slow (S) acoustic waves at the frequency of 100 kHz.}
  \label{tab:noise_levels}
 \end{center}
\end{table*}

Relatively low values are obtained for each case, which is due to the fact that the estimation has been made for a narrow frequency range (100 - 101 kHz). 
The higher noise levels estimated for a slow-wave-dominated free-stream are due to the lower wall-response levels observed in general for slow acoustic waves in all the considered numerical cases, compared to the response to fast acoustic waves. If the comparison is restricted to the higher Mach number cases (Mach 7.3 and Mach 6), the values listed in table \ref{tab:noise_levels} indicate higher free-stream noise levels in the HEG facility at the highest Mach number, compared to the RWG facility. The estimation for the Mach 3 case in the RWG facility (Case 5) reveals a further decrease of the noise level relative to fast acoustic waves, and, in contrast, a significant increase of the noise level for slow acoustic waves, compared to the higher Mach number cases. The higher value reached for the Mach 3 case with slow acoustic waves is mainly due to the very low value of the wall response predicted by the DNS (see figure \ref{fig:RWG_M3_spectrum}), lower by almost 2 orders of magnitude compared to the higher Mach number cases, which is not followed by a similar reduction in the experimental pressure values (as seen in table \ref{tab:noise_levels}). With reference to the experimental values in table \ref{tab:noise_levels}, it should be noticed that, while moving from Mach 7 to Mach 6 a reduction by about 1/3 of the measured value is registered, a reduction slightly lower than 1/6 is observed when moving from Mach 6 to Mach 3, which appears substantially smaller than what would be expected considering the strong reduction in the Mach number. 
This suggests that the RWG facility in the low Mach number case might either have been affected by other sources of noise, apart from the acoustic source (e.g. entropy spottiness or vorticity), which would have increased the overall disturbance level, and/or that the significantly higher Mach angle (compared to the higher Mach number cases) of the inclined waves radiated from the turbulent boundary layer at the nozzle walls has led to an increase of the disturbance amplitude on the probe surface. 
The estimated value for the Mach 3 case with a slow-wave dominated free-stream ($1.26\times10^{-3}$) is in relatively good agreement with the free-stream pressure fluctuation level of $3.96\times10^{-3}$ obtained by Duan et al. \cite{Duan2014a} in a DNS study on the acoustic noise generated by a turbulent boundary layer over a flat plate at Mach 2.5 (thus comparable with our Mach 3 case). 
In general, the higher values associated with slow waves can be considered more realistic than the very small values relative to fast acoustic waves. This can be seen as addition empirical evidence that the slow acoustic waves are indeed the dominant acoustic disturbances in hypersonic wind tunnels (Duan et al.\cite{Duan2014a}).

\section{Summary}

A wedge-shaped probe was introduced to quantify free-stream disturbances in a wide range of hypersonic facilities covering facilities with harsh test environments such as shock tunnels. The probe was designed allowing easy implementation and exchange between different facilities. Using the probe at an angle of attack allows the signal-to-noise ratio to be adjusted to different test environments while still providing a 2D-flow field on the probe to not compromise the measurements. The slender probe design avoids the complex subsonic flow field typically found behind blunt probes and thus simplifies the transfer function of free-stream disturbances entering the boundary layer across the oblique shock. Furthermore, the probe design sufficiently protected the flush mounted instrumentation against thermal loads and particle impact while at the same time a good high frequency response is ensured by avoiding resonance and damping effects due to protective cavities. \\
In the scope of the present study comparative tests were conducted in three hypersonic facilities, i.e. in the DLR High Enthalpy Shock Tunnel G\"ottingen (HEG) at Mach 7.4, in the DNW-RWG Ludwieg tube at Mach 3 and Mach 6 and in the TU Braunschweig Ludwieg tube (HLB) at Mach 6. KULITE\textsuperscript{\textregistered} and PCB\textsuperscript{\textregistered} pressure transducers were found to be particularly useful and reliable although the latter needed special treatment to avoid mechanical vibration effects. Thin film transducers and thermocouples were found to provide very low signal-to-noise ratios in cold hypersonic flows, making the application in Ludwieg tube facilities pointless. The ALTP transducer showed good signal to noise ratio in all three facilities. However, the transducers sensitivity is subject to strong uncertainties. Therefore, the ALTP readings were excluded in the present analysis. The tests in the two Mach 6 wind tunnels independently revealed that in the low frequency range the spectral energy decreases with increasing unit Reynolds number whereas the trend reverses in the high frequency range. The frequency at which the reversal appears was found to be approximately \unit{50}{kHz} in RWG and approximately \unit{100}{kHz} in HLB, resulting in slightly lower RMS pressure readings in RWG at higher Reynolds numbers. Since both facilities have identical nozzle exit diameters and similar free-stream Mach numbers nozzle scaling effects are not expected to cause this difference. The normalized surface pressure RMS over a wide Reynolds number range were determined to be approximately \unit{1-1.6}{\%} up to \unit{50}{kHz}, \unit{0.25-0.4}{\%} between \unit{50-100}{kHz}, \unit{0.15-0.3}{\%} between \unit{100-200}{kHz} and below \unit{0.2}{\%} between \unit{200-300}{kHz}. Furthermore, pitot pressure fluctuations were compared to recent data obtained in various hypersonic facilities revealing a good agreement. It could be shown that the results obtained with the wedge probe can be converted into corresponding pitot pressure fluctuations. The converted data were found to agree well at Mach numbers 7.4 and 6. This observation supports the intention of the present study to provide a probe to be able to access test environments that cannot be investigated by means of pitot probes or hot wires. \\
Complementary to the experimental study, numerical simulations were performed to study the receptivity to fast and slow acoustic waves at the experimental flow conditions. Transfer functions for the pressure fluctuations were computed for different imposed frequencies and used to estimate the free-stream noise levels for a range of flow conditions for fast and slow acoustic waves, which were considered to be the principal free-stream disturbances. The receptivity to fast acoustic waves was observed to be characterized by an early amplification of the induced fast mode, followed by a region, dependent on the Mach number and on the frequency, of modulation between the fast mode and the forcing fast acoustic waves. In contrast, in the receptivity to slow acoustic waves, the response showed an initial decay region close to the leading edge, followed by a region of substantially constant amplitude level further downstream. Overall, the leading-edge receptivity to fast acoustic waves was found to be higher than the receptivity to slow acoustic waves, at all  Mach numbers and for all considered frequencies. \\
For all the Mach numbers, the simulated frequency spectra of the wall pressure showed a flat profile compared to the decreasing trend of the experimental frequency spectra, for a constant amplitude of the free-stream disturbances imposed at all the frequencies, demonstrating that, for the considered cases, the receptivity process has only a weak effect on the shape of the frequency spectra of the wall response. As a result, the shape of the estimated free-stream noise level profiles, obtained through a combination of the experimental wall pressure fluctuation levels and the numerical free-stream-to-wall transfer functions, appeared substantially unaltered relative to the shape of the experimental wall pressure levels. 
%%% Updated conclusion with new noise levels %%%
An estimation of the free-stream noise levels for a frequency of 100 kHz, as a combination of the numerical transfer functions and the experimental RMS levels in a frequency range of 100-101 kHz, revealed noise levels on the order of $10^{-6}$ and $10^{-5}$ if solely fast acoustic waves are considered. Considering only slow acoustic waves higher values were found, ranging from the order of $10^{-5}$, for the Mach 6 case, to $10^{-3}$ for the Mach 3 case. For the higher Mach number cases (Mach 7.3 and Mach 6) it was found that the HEG facility is noisier than the RWG. The predicted noise levels for fast waves were found to decrease at decreasing Mach numbers, while for slow acoustic waves, in contrast, the highest noise level was found for the Mach 3 case (1.26$\times 10^{-3}$) in the RWG facility. A good agreement was found for the latter value with independent DNS results available in the literature. It is thus likely that the slow-wave cases provide in general more realistic values of the free-stream noise levels compared to the fast-wave cases.

\begin{acknowledgements}
The present study was supported by an ESA funded Technology Research Project (ESA-Contract number $4200022793/09/NL/CP$). The received support is gratefully acknowledged. Furthermore, the authors wish to acknowledge the assistance of the HEG staff, in particular Jan Martinez Schramm, Ingo Schwendtke and Uwe Frenzel.
\end{acknowledgements}

\bibliography{Transition}

\end{document}